\documentclass[10pt]{IEEEtran}

\usepackage{graphicx}
\usepackage[colorlinks,linkcolor=black,anchorcolor=black,citecolor=black]{hyperref}
\usepackage{lineno,hyperref}
\modulolinenumbers[5]
\usepackage[cmex10]{amsmath}
\usepackage{amsfonts}

\usepackage{color}
\usepackage{algorithm}
\usepackage{algorithmic}
\usepackage{array}
\usepackage{multirow}
\usepackage{booktabs}
\usepackage{bm}

\allowdisplaybreaks[4]
\usepackage{amsmath}
\usepackage{underscore}
\usepackage{stfloats}
\usepackage{float}
\usepackage{lipsum}
\usepackage{cite}
\usepackage{subfigure}
\usepackage{adjustbox}
\usepackage{amssymb}
\usepackage{amsmath}

\makeatletter
\newenvironment{breakablealgorithm}
{
	\begin{center}
		\refstepcounter{algorithm}
		\hrule height.8pt depth0pt \kern2pt
		\renewcommand{\caption}[2][\relax]{
			{\raggedright\textbf{\ALG@name~\thealgorithm} ##2\par}%
			\ifx\relax##1\relax 
			\addcontentsline{loa}{algorithm}{\protect\Gammamberline{\thealgorithm}##2}%
			\else 
			\addcontentsline{loa}{algorithm}{\protect\Gammamberline{\thealgorithm}##1}%
			\fi
			\kern2pt\hrule\kern2pt
		}
	}{
		\kern2pt\hrule\relax
	\end{center}
}

\begin{document}

\title{Adaptive Resource Allocation for Semantic Communication Networks}

\author{Lingyi Wang,
Wei Wu,~\IEEEmembership{Member,~IEEE,}
Fuhui Zhou,~\IEEEmembership{Senior Member,~IEEE,}\newline
Zhaohui Yang,~\IEEEmembership{Senior Member,~IEEE,}
Zhijin Qin,~\IEEEmembership{Senior Member,~IEEE}

\thanks{This work was supported by the National Key Research and Development Program of China under Grant 2020YFB1807602;
and the National Natural Science Foundation of China under Grant 62271267.}
\thanks{Lingyi Wang is with the College of Science,
Nanjing University of Posts and Telecommunications, Nanjing, 210003, China
(e-mail: lingyiwang@njupt.edu.cn).}
\thanks{
Wei Wu is with the College of Communication and Information Engineering,
Nanjing University of Posts and Telecommunications, Nanjing, 210003, China
(e-mail: weiwu@njupt.edu.cn).}
\thanks{Fuhui Zhou is with the College of Electronic and Information Engineering, 
Nanjing University of Aeronautics and Astronautics, Nanjing, 210000, China
(e-mail: zhoufuhui@ieee.org).}
\thanks{Zhaohui Yang is with Zhejiang Lab, Hangzhou 311121, China, also with
the College of Information Science and Electronic Engineering, Zhejiang
University, Hangzhou, Zhejiang 310027, China, and also with the Zhejiang Provincial Key Laboratory of Information Processing, Communication and Networking (IPCAN), Hangzhou, Zhejiang 310007, China 
(e-mail: yang_zhaohui@zju.edu.cn).}
\thanks{Zhijin Qin is with the Department of Electronic Engineering, Tsinghua University, Beijing 100084, China.
(e-mail: qinzhijin@tsinghua.edu.cn).}
}

\maketitle
\begin{abstract}
  Semantic communication, recognized as a promising technology for future intelligent applications, has received widespread research attention. 
  Despite the potential of semantic communication to enhance transmission reliability, especially in low signal-to-noise (SNR) environments, 
  the critical issue of resource allocation and compatibility in the dynamic wireless environment remains largely unexplored. 
  In this paper, we propose an adaptive semantic resource allocation paradigm with semantic-bit quantization (SBQ) compatibly for existing wireless communications, where the inaccurate environment perception introduced by the additional mapping relationship between semantic metrics and transmission metrics is solved.
  In order to investigate the performance of semantic communication networks, the quality of service for semantic communication (SC-QoS), including the semantic quantization efficiency (SQE) and transmission latency, is proposed for the first time.  
  A problem of maximizing the overall effective SC-QoS is formulated by jointly optimizing the transmit beamforming of the base station, the bits for semantic representation, the subchannel assignment, and the bandwidth resource allocation.
  To address the non-convex formulated problem, an intelligent resource allocation scheme is proposed based on a hybrid deep reinforcement learning (DRL) algorithm,
  where the intelligent agent can perceive both semantic tasks and dynamic wireless environments.
  Simulation results demonstrate that our design can effectively combat semantic noise and achieve superior performance in wireless communications compared to several benchmark schemes.
  Furthermore, compared to mapping-guided paradigm based resource allocation schemes, our proposed adaptive scheme can achieve up to $\mathrm{13\%}$ performance improvement in terms of SC-QoS.
\end{abstract}

\begin{IEEEkeywords}
Semantic communication, quality of service, resource allocation, deep reinforcement learning, semantic-bit quantization. 
\end{IEEEkeywords}

\IEEEpeerreviewmaketitle

\section{Introduction}
The rapid development of intelligent devices and the exponential growth of wireless data traffic are driving communication networks toward high-efficiency and intelligence\cite{LU2020100158}. 
Effectively handling ultra-large-scale connections and enabling intelligent applications within limited spectrum resources pose a significant challenge for the development of the sixth-generation (6G) wireless communication systems \cite{9606720,8808168}. 
The conventional communication paradigm has predominantly emphasized accurate symbol, but the surge in intelligent radio services has rendered the network capacity insufficient to accommodate the increasing demand for high-volume traffic \cite{qin2021semantic}. 
To address this issue, the semantic communication paradigm has been proposed, leveraging machine learning techniques to extract task-related semantic features from raw data, with the objective of completing tasks \cite{9530497}. 
The current work \cite{9398576,9954153,9837870,9953076,9763856,10122232} has demonstrated that semantic communication can drastically alleviate network congestion and enable robust transmission in low signal-to-noise (SNR). 
Majority of work \cite{9398576,9954153,9837870,9953076} focuses on the semantic coding with the end-to-end (E2E) framework and the joint source-channel coding (JSCC) design, while semantic resource allocation remains being explored.

Semantic resource allocation currently faces several challengings:
\textit{1) How to optimize the resource allocation accurately at a semantic level? 
2) How to be compatible with existing wireless communication hardware devices?
3) How to conceptualize semantic resource allocation and machine learning-based semantic coding as a whole framework?}
Although the current resource allocation schemes proposed in \cite{9763856} and \cite{10122232} have attempted to optimize the resource allocation at a semantic level while additionally pre-trained mappings are demanded to guide the network resource allocation, where the relationships between transmission metrics and semantic metrics are pre-recorded. 
This mapping-guided semantic resource allocation paradigm is limited in its ability to address the time-varying wireless environment and make it difficult to allocate network resource accurately.
Hence, in this paper, we aim to consider a non-guided adaptive paradigm, eliminating such static mappings, and directly guiding semantic resource allocation based on dynamic communication characteristics and semantic task characteristics.

For challenge 2, the codebook-based semantic knowledge base \cite{10101778} achieves the discretization of continuous semantic information, which allows semantic information to be transmitted over bits in wireless communications.
However, since this quantification is strongly task-related, a more generalized discretization approach is needed.
Moreover, orthogonal frequency division multiple (OFDM) can avoid inter-subchannel interference and improve the spectrum efficiency and achieve better transmission quality with multiple receivers \cite{9973061,10107714}.
Hence, the OFDM networks with multiplicative and non-differentiable channels are considered for wireless communications in this paper.
Regarding to challenge 3, it is difficult for traditional mathematical methods to support real-time computation and work in collaboration with AI-driven semantic coding, especially for complex large-scale intelligent communication scenarios.
Recently, deep reinforcement learning (DRL) is considered as an intelligent optimization scheme to quickly solve complex problems and achieve excellent real-time performance with intelligence \cite{9973061,wang2023intelligent}.
Therefore, DRL serves as the optimal solution for intelligent semantic resource allocation.

Noting the tricky challenges above, this paper investigates an adaptive resource allocation scheme with novel semantic-bit quantization (SBQ) and intelligent offset compensation for the non-guided semantic communication paradigm.
SBQ provides an effective solution for encoding continuous semantic features into bits suitable for transmission in wireless communication systems, 
while offset compensation can effectively combat the semantic noise caused by SBQ loss and bit error rate (BER) in wireless channels, particularly at the low SNR.
Furthermore, AI-based resource allocation methods should synergistically work with AI-driven semantic communication to construct forward-looking intelligent communication networks.
However, very few investigations have been done. 
Our main contributions are summarized as follows. 
\begin{itemize}
  \item  It is the first time that an adaptive resource allocation scheme is proposed for wireless semantic communication with SBQ and intelligent compensators.
  The SBQ facilitates the coding between semantics and bits while the compensator can combat the semantic noise caused by SBQ loss and BER.
  \item  The QoS of semantic communication (SC-QoS) is initially defined based on the efficiency of semantic quantization (SQE) and the transmission latency,
  which can effectively evaluate system performance and user experience with finite network resources.
  \item  The transmit beamforming of the base station, the bits for semantic representation, the bandwidth allocation, and subchannel assignment are jointly optimized to maximize the effective SC-QoS.
  \item  To address formulated intricate problems rapidly and propel intelligent semantic resource allocation schemes, 
  a hybrid DRL-based dynamic resource allocation scheme is designed to address the hybrid spaces with coupled variables and integer programming.
  \item  Simulation results demonstrate that our proposed resource allocation schemes achieve a improvement of $\mathrm{13\%}$ in terms of SC-QoS compared to the existing mapping-guided schemes for wireless semantic communications.
\end{itemize}

The remainder of this paper is organized as follows. 
Section II introduces the related works.
Section III presents the semantic communication system including semantic coding, performance metrics and problem formulation. 
Section IV proposes our intelligent semantic resource allocation scheme. 
The simulation results are given in Section V. 
Lastly, Section VI draws the conclusion for this paper.

\textit{Notation:} Scalars are represented by using italics, while boldface letters denote vectors and matrices. 
$\mathcal{C} \mathcal{N}\left(\mu, \sigma^2\right)$ represents a random variable that follows a circularly symmetric complex Gaussian distribution with the mean of $\mu$ and the variance of $\sigma^2$. 
The set of $n \times m$ real matrices is denoted by $\mathbb{R}^{n \times m}$. 
For any vector $\mathbf{u}$, its transpose is denoted by $\mathbf{u}^T$, and $\operatorname{diag}(\mathbf{u})$ denotes a diagonal matrix with diagonal elements corresponding to the elements of $\mathbf{u}$. 
The space of complex-valued matrices of size $x \times y$ is denoted by $\mathbb{C}^{x \times y}$, and the mathematical expectation is represented by $\mathbb{E}[\cdot]$. 
The set of elements is denoted by $\{\cdot\}$, while $\arg \underset{x}{{\max}}g(x)$ represents the value of $x$ that maximizes the function $g(x)$.

\vspace{-0.15cm}
\section{Related Work}
\subsection{\textit{Semantic Coding for Semantic Communications}}
As pointed out in \cite{10101778}, semantic coding for semantic communication can be divided into two paradigms, 
namely, the full-resolution constellation enabled and the limited-resolution constellation enabled.
Majority of the existing studies directly encode the discrete source data into continuous channel symbols \cite{9398576,9832831,9450827,9796572,9830752,9959884}, which belongs to the first paradigm where the position of constellation points is considered to be unrestricted within the constellation.
In \cite{9398576}, the authors proposed an E2E text semantic coding framework with BERT-based similarity metric at the semantic level.
The data modality was expanded to speech coding \cite{9450827}, image semantic coding \cite{9796572,9830752,9959884} and video semantic coding \cite{9837870}.
A DRL-based scheme was proposed in \cite{9796572} to effectively capture the important semantic blocks for the back-end classifier, where the channel gain and the semantic information were jointly considered.
In \cite{9830752}, the authors explored a unified semantic communication framework for multi-user tasks.
The authors in \cite{9959884} proposed a personalized semantic encoder considering the interests of users.

However, the implementation of the full-resolution constellation in wireless communication systems poses significant challenges, leading to the proposition of the second paradigm\cite{10101778,zhang2022unified}.
In the limited-resolution constellation enabled semantic coding paradigm, the constellation is characterized by limited points, and the semantics is encoded into bits for the physical wireless channel.
The authors in \cite{10101778} and \cite{zhang2022unified} used a codebook discretize semantic information.
Specifically, the learnable codebook consists of an $N$-dimensional approximate orthogonal vector set, and the index of the vector with the closest distance to the semantic information is selected as the semantic representation, which is then transformed into bit information.
While the utilization of codebooks represents a significant step forward in implementing the second paradigm, training an orthogonal set of high-dimensional vectors poses challenges in terms of learning complexity. 
Moreover, there is a noticeable loss of accuracy in the semantic quantization and physical transmission when relying on vector indexs, leading to substantial semantic noise. 
As a consequence, a unified quantization method needs further investigation.
Furthermore, the aforementioned works \cite{9398576,9832831,9450827,9796572,9830752,9959884} have mainly concentrated on studying the semantic coding framework. 
It is equally critical to efficiently allocate resources to alleviate spectrum scarcity in future communication networks characterized by ultra-large scale connections and ultra-high traffic volumes.

\vspace{-0.35cm}
\subsection{\textit{Resource Allocation for Semantic Communication}} 
In view of the transmission characteristics of the two semantic coding paradigms, resource allocation for semantic communication networks are divided into two corresponding paradigms.
The first one is the mapping-guided paradigm, where a static mapping is desired to optimize resource allocation at the semantic level for wireless communications based on the full-resolution constellation enabled semantic coding paradigm\cite{9763856,10001594,liu2022adaptable,10122232}.
The second one is the non-guided adaptive paradigm, where no mapping is required, and dynamic wireless environment can be directly awared based on the limit-resolution constellation enabled semantic coding paradigm.

Based on the DeepSC, the authors in \cite{9763856} defined the semantic spectral efficiency (S-SE) for the first time.
As mentioned in \cite{9763856}, the semantic similarity performance was deposited in a joint mapping of the semantic symbols and SNR.
Therefore, static mappings introduce the evaluation accuracy loss, related to the stored SNR accuracy, and additional storage costs are required for higher estimation accuracy.
The authors in \cite{10001594} further considered the quality-of-experience based on \cite{9763856}.
Different from the discrete mappings in \cite{9763856,10001594}, a curve-fitting method was employed to further formulate a continuous mapping relationship \cite{10122232}.
Whilst the works \cite{9763856,10001594,10122232} have considered resource allocation schemes for semantic communication, 
semantic coding and resource allocation remain independent processes, and the intermediate mappings are necessary to serve as a guidance of resource allocation.
It is worth noting that mapping-guided resource allocation schemes suffer from precision loss and failure to sense the real-time wireless communication environment.
This is because the mapping-guided paradigm is built upon the full-resolution constellation enabled semantic coding with restricted application in wireless communications.
To address the aforementioned challenges, the second paradigm using the limited-resolution constellation enabled semantic coding without static mappings is proposed. 
To the best knowledge of the authors, the non-guided adaptive paradigm for resource allocation has not been studied. 
Hence, this paper investigates an adaptive resource allocation for semantic communication networks, considering the dynamic channel characteristics and providing efficient semantic communication service.
 
\vspace{-0.3cm}
\subsection{\textit{DRL-based Intelligent Resource Allocation}}
DRL has gained significant popularity in traditional communication systems due to its aptitude for swiftly suboptimal solutions to intricate non-convex problems.
In \cite{9031419,9973061,9046301,wang2023intelligent}, DRL-based resource allocation schemes were considered in spectrum sharing networks.
The authors of \cite{9973061} proposed novel hybrid DRL approaches to address the discrete action space and the continuous action space.
Based on the distributed proximal policy optimization approach\cite{9046301}, the wireless communication system was capable of adaptively adjusting the transmit power.
Based on the hybrid DRL approaches, the work in \cite{9973061} and \cite{wang2023intelligent} further maximized the secure rate of the secondary network under the presence of eavesdroppers.
The preceding studies have showcased the efficacy and real-time proficiency of DRL in tackling intricate non-convex challenges, 
thereby rendering it highly suitable for large-scale communication systems. 
In the context of uplink non-rrthogonal multiple access communications, the authors in \cite{9381701} have proposed a DRL-based resource allocation strategy, which was superior to the conventional orthogonal multiple access-based Internet of things (IoT) networks in system throughput.
Moreover, DRL-enabled dynamic resource allocation schemes exhibited robust computing capabilities and minimal latency, exemplifying significant performance improvement relative to the benchmark schemes for mobile edge computing networks\cite{9481307,9435782,9759989}.
Nevertheless, little attention has been paid to explore the DRL-based semantic resource allocation\cite{10122232}, which holds tremendous potential in furnishing semantic communication systems with intelligent and large-scale deployment support.

\vspace{-0.1cm}
\section{Semantic Communication Framework wiht Semantic-Bit Quantization}
In this section, a novel semantic communication framework with a novel SBQ is proposed.
A downlink communication network with multiple users is considered.

\vspace{-0.3cm}
\subsection{Semantic Coding}
As shown in Fig. 1, the semantic encoder extracts semantic features from the raw data $\mathbf{d}_{u}[i]$ at the base station, and the semantic decoder reconstructs the raw data from the received information $\widehat{\mathbf{d}}_{u}[i]$ at the user side.
Both the semantic encoder and semantic decoder are constructed based on the Transformer \cite{vaswani2017attention},
since Transfomer demonstrated superior performance compared to benchmark algorithmic frameworks such as RNN and CNN on natural language tasks\cite{devlin2018bert}.
The control center of the base station is treated as an intelligent agent that can provide resource allocation service by the dynamical transmit beamforming, subchannel assignment, bandwidth allocation and SBQ.
Each user is equipped with a single antenna, while the base station is equipped with $M$ antennas. 
Let $\mathcal{U} = \{1, \dots ,U\}$ and $\mathcal{C} = \{1, \dots ,C\}$ respectively represent the sets of users and subchannels.
It is assumed that $C$ is larger than $U$, in order that each user can occupy one subchannel.

\begin{figure*}
  \centering
  \includegraphics[scale=0.4]{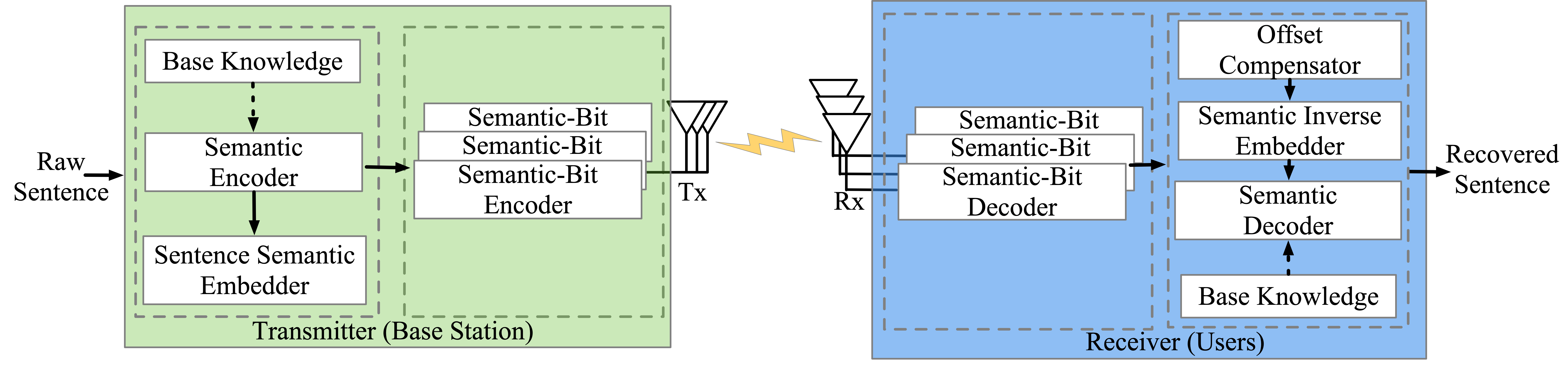}
  \vspace{-0.1cm}
  \caption{The framework of our proposed semantic communication.}
  \vspace{-0.5cm}
\end{figure*}

At the base station side, the semantic features are extracted from the $i$-th raw data $\mathbf{d}_{u}[i]$ at the $u$-th user. 
The $i$-th sentence data transmitted to the $u$-th user can be represented by $\mathbf{d}_{u}[i]=\left[{w_{u}[i]}^1, w_{u}[i]^2, \ldots, w_{u}[i]^l, \ldots, w_{u}[i]^{L_{u}[i]}\right]$, where $L_{u}[i]$ denotes the length of the sentence and $w_{u}[i]^l$ denotes the word at position $l$ in the sentence $\mathbf{d}_{u}[i]$, with $l \in [1, L_{u}[i]]$.
Hence, the representation of the sentence semantic features (semantic information) of the $i$-th raw data for the $u$-th user can be denoted by
\begin{equation}
  \mathbf{x}_{u}[i]= \mathcal{F}(\mathbf{d}_{u}[i];{\beta}),
\end{equation}
where ${\beta}$ is the parameter of the semantic encoder $\mathcal{F}(\cdot)$.
The semantic information is input into the sentence embedder for semantic compression, and the semantic embedding information is obtained by
\begin{equation}
  \mathbf{e}_{u}[i] = \mathcal{E}(\mathbf{x}_{u}[i];{\phi}),
\end{equation}
where ${\phi}$ is the parameter of the sentence semantic embedder $\mathcal{E}(\cdot)$.
Different from $\mathbf{y}=\mathbf{H}\mathbf{x}+\mathbf{n}$, $\mathbf{n} \sim \mathcal{C} \mathcal{N}\left(\mathbf{0}, \sigma^2 \mathbf{I}\right)$, proposed in \cite{9832831,9450827,9796572,9830752,9959884}, the channel coding was employed to directly map continuous semantic features into channel symbols.
Nevertheless, the implementation of this concept faces challenges to be applied compatibly in the existing wireless communication systems \cite{10101778}.
In this paper, a semantically efficient quantization scheme is designed for converting semantics into bits, targeting the wireless semantic transmission for the current digital communication.
The processing of the wireless semantic transmission is written as $\mathcal{T}(\cdot)$, including SBQ, modulation, channel transmission, and demodulation.

At the user side, the received sentence  information $\widehat{\mathbf{e}}_{u}[i] = \mathcal{T}(\mathbf{e}_{u}[i])$ is input into the semantic inverse embedder to obtain the semantic features $\widehat{\mathbf{x}}_{u}[i]$, which is denoted by
\begin{equation}
  \widehat{\mathbf{x}}_{u}[i] = \mathcal{E}^{-1}(\widehat{\mathbf{e}}_{u}[i];{\eta}),
\end{equation}
where ${\eta}$ is the parameter set of the semantic inverse embedder $\mathcal{E}^{-1}(\cdot)$.
Finally, the sentence is reconstructed with the semantic decoder based on the recovered sentence semantic embedding $\widehat{\mathbf{x}}_{u}[i]$.
The processing of the semantic decoding is represented by 
\begin{equation}
  \widehat{\mathbf{d}}_{u}[i] = \mathcal{F}^{-1}(\widehat{\mathbf{x}}_{u}[i];{\kappa}),
\end{equation}
where ${\kappa}$ is the parameter set of the semantic decoder $\mathcal{F}^{-1}(\cdot)$.

To assess the discrepancy between the original and recovered sentence throughout the training process, 
the loss function based on the cross-entropy (CE) is conducted \cite{9398576}, which is denoted by
\begin{equation}\label{ST}
  \begin{aligned}
  &\mathcal{L}^{\mathrm{SE}}(\{\mathbf{d}_{u}[i]\}, \{\widehat{\mathbf{d}}_{u}[i]\} ; {\beta}, {\phi}, {\eta}, {\kappa})= -\sum_{u}^{}\sum_{l_d}^{} q\left(w_{u}[i]^{l_d}\right) \\
  &\log \left(p\left(w_{u}[i]^{l_d}\right)\right)+\left(1-q\left(w_{u}[i]^{l_d}\right)\right) \log \left(1-p\left(w_{u}[i]^{l_d}\right)\right).
  \end{aligned}
\end{equation}
In the given context, $q\left(w_{u}[i]^{l_d}\right)$ represents the actual probability of the $l_d$-th word, appearing in the raw sentence $\mathbf{d}_u[i]$, and $p\left(w_{u}[i]^{l_d}\right)$ denotes the predicted probability of the $l_d$-th word, appearing in the recovered sentence $\widehat{\mathbf{d}}_u[i]$.
The details of training the semantic coding are summarized in $\textbf{Algorithm}$ 1.
\begin{breakablealgorithm}
  \caption{Train Semantic Coding Model.}
  \label{alg:alg1}
  \begin{algorithmic}[1]
    \STATE \textbf{Input:} The knowledge vocabulary $V$.
    \STATE \textbf{Transmitter:} 
    \STATE \hspace{\algorithmicindent} Data source $V$ $\rightarrow$ $\{\mathbf{d}_u[i]\}$.
    \STATE \hspace{\algorithmicindent} $\mathcal{F}(\mathbf{d}_u[i];{\beta}) \rightarrow \{\mathbf{x}_u[i]\}$.
    \STATE \hspace{\algorithmicindent} $\mathcal{E}(\mathbf{x}_u[i];{\phi}) \rightarrow \{\mathbf{e}_u[i]\}$.    
    \STATE \textbf{Physical Channel:} 
    \STATE \hspace{\algorithmicindent} Assume error-free bits.
    \STATE \textbf{Receivers:} 
    \STATE \hspace{\algorithmicindent} $\mathcal{E}^{-1}(\mathbf{e}_u[i];{\eta}) \rightarrow \{\widehat{\mathbf{x}}_u[i]\}$. 
    \STATE \hspace{\algorithmicindent} $\mathcal{F}^{-1}(\widehat{\mathbf{x}}_u[i];{\kappa}) \rightarrow \{\widehat{\mathbf{d}}_u[i]\}$.   
    \STATE \hspace{\algorithmicindent} Computer the loss $\mathcal{L}^{\mathrm{SE}}$ given by (\ref{ST}).
    \STATE \hspace{\algorithmicindent} Update ${\beta},{\phi},{\eta},{\kappa}$ with gradient descent.
    \STATE \textbf{Output:} The trained parameter sets $\{{\beta},{\phi},{\eta},{\kappa}\}$.
  \end{algorithmic}
\end{breakablealgorithm}

\vspace{-0.2cm}
\subsection{Proposed Semantic-Bit Quantization for Digital Communication}
In this part, a novel semantic communication structure by using SBQ scheme with low complexity is proposed.
\begin{figure}
  \centering
  \setlength{\subfigcapskip}{-0.3cm}
  \subfigure[The distribution of semantic features.]{
    \begin{minipage}{8cm}
    \includegraphics[width=\textwidth]{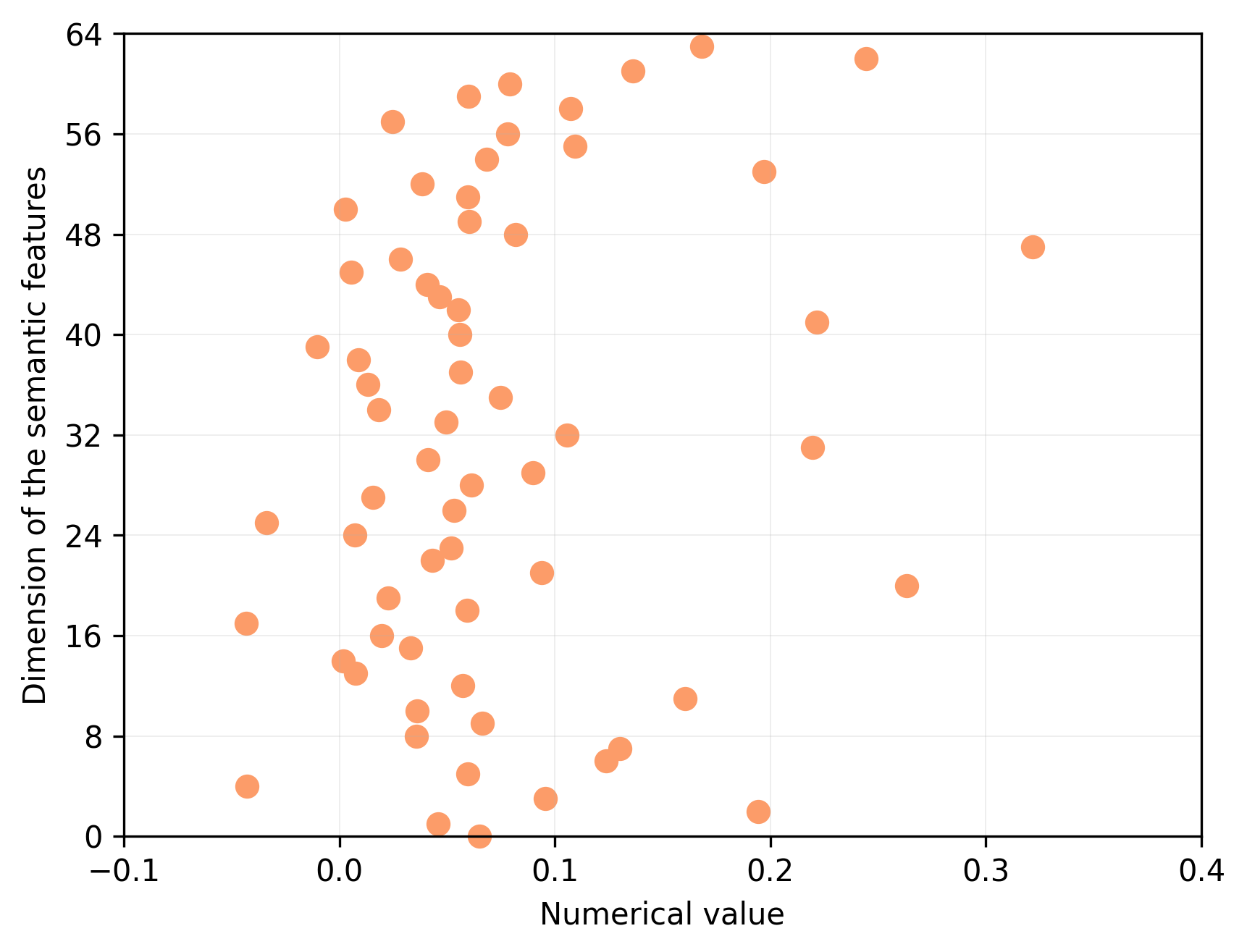} \\
    \end{minipage}}
  \subfigure[Comparison of different quantization threshold.]{
    \begin{minipage}{8cm}
      \includegraphics[width=\textwidth]{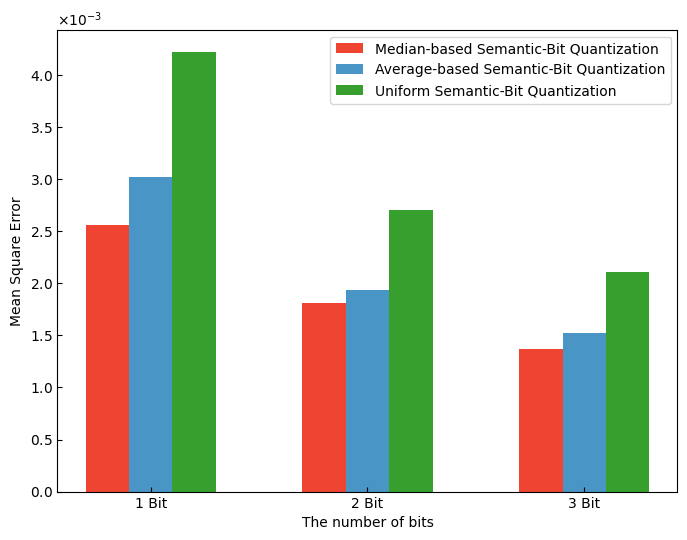} \\
    \end{minipage}}
  \caption{Our proposed SBQ scheme.} 
  \vspace{-0.5cm}
\end{figure}
\subsubsection{The Design of SBQ}
As depicted in Fig. 2(a), the mathematical distribution of semantic features is non-uniform. Hence, the conventional uniform quantization approach makes it difficult to directly and effectively transform float-based semantic features into bit representations.
To promote the compatible deployment of wireless semantic communication, we propose a novel semantically efficient SBQ scheme.
The float-based sentence embeddings $\{\mathbf{e}_{u}[i]\}$ need to be transformed into the bit-based embedding information $\{\mathbf{e}_{u}^{\mathrm{bit}}[i]\}$ in the digital communication systems.
In this paper, the considered knowledge base is the European Parliament \cite{9398576}.
The learnable semantic quantization threshold $p^{\mathrm{th}}$, $p^{\mathrm{th}}>0$, related to the specific semantic knowledge base and semantic coding networks, is proposed to evaluate an upper bound on the value of high-density semantic features.
Let $\mathcal{N} = \{1, 2,\dots,n,\dots ,N\}$ denote the SBQ set, where the $n$-th semantic-bit codec encodes and decodes sentence embedding position with $n$ bits.
$p^{\mathrm{float}}$ denotes the mathematical value of the semantic feature, $p^{\mathrm{max}}_n$ denotes the maximum value of positive definite points that can be represented by $n$ bits. 

Uniform quantization with $n$ bit is applied for values falling within the specified high-density ranges, $(-\infty,p^{\mathrm{th}}]$, which is represented by
\begin{equation}
  p^{\mathrm{int}}_n = \mathrm{clamp}(0, p^{\mathrm{max}}_{n-1}, \mathrm{round}(\frac{p^{\mathrm{float}}}{p^{\mathrm{th}}}p^{\mathrm{max}}_{n-1})),
\end{equation}
where the function $\mathrm{clamp}(x_{\min}, x_{\max}, x)$ means limitation on $x$ within a specified range, and the function $\mathrm{round}(x)$ means the nearest integer to $x$. 
The reconstructed float-based semantic feature $p^{\mathrm{float}}$ is denoted by
\begin{equation}
  \widehat{p}^{\mathrm{float}} =\frac{2 p^{\mathrm{int}}+1}{2 p^{\mathrm{max}}_{n-1}}p^{\mathrm{th}}.
\end{equation}

However, there are still some remaining semantic feature outliers, $(p^{\mathrm{th}},\infty)$, and in order to characterize as many of these features as possible, non-uniform quantization is used, which is denoted by
\begin{equation}
  p^{\mathrm{int}}_n = \mathrm{clamp}(p^{\mathrm{max}}_{n-1}, p^{\mathrm{max}}_{n}, \mathrm{round}(\frac{p^{\mathrm{float}}-p^{\mathrm{th}}}{p^{\mathrm{th}}})+p^{\mathrm{max}}_{n-1}).
\end{equation}
The recovered float-based semantic feature $p^{\mathrm{float}}$ is denoted by
\begin{equation}
  \widehat{p}^{\mathrm{float}}_n =(p^{\mathrm{int}}-p^{\mathrm{max}}_{n-1}+1)p^{\mathrm{th}}.
\end{equation}
Based on the above approach, each semantic feature in $\mathbf{e}_{u}[i]$ is encoded into discrete bits in $\mathbf{e}_{u}^{\mathrm{bit}}[i]$, then modulated, transmitted with OFDM, and demodulated.

The semantic quantization threshold $p^{\mathrm{th}}$ is set to the average and median value of the semantic features generated by the semantic encoder, comparing with the uniform quantization scheme in mean square error.
As depicted in Fig. 2(b), it is evident that our proposed hybrid uniform-nonuniform quantization method can efficiently represent semantic compared to the uniform quantization method due to the nonuniform distribution of the semantic features.
Moreover, the median value is a more suitable choice for serving as the quantization threshold in our specific knowledge base. 

\subsubsection{Intelligent Compensator}
Despite the efficient encoding of semantic information into bits, the quantization process encounters considerable semantic noise, leading to a notable loss of semantic information precision.
Moreover, the signal experiences distortions during transmission due to the instability and noise of the physical channel, particularly at low SNRs, further impacting the semantic accuracy.
Hence, in this context, two main challenges arise: (1) \textit{How to effectively eliminate the semantic noise caused by SBQ?} (2) \textit{How to effectively address the semantic noise caused by error bits with dynamic channel characteristics?} 

To effectively combat the semantic noise caused by SBQ loss and channel characteristics, 
the intelligent offset compensator is designed to adjust the positional offset of the recovered sentence embedding $\mathring{\mathbf{e}}_{u}[i]$ at the receiver,
denoted by
\begin{equation}
  \widehat{\mathbf{e}}_{u}[i] = \mathcal{P}(\mathring{\mathbf{e}}_{u}[i];{\epsilon}),
\end{equation}
where ${\epsilon}$ is the parameter set of the semantic offset compensator $\mathcal{P}(\cdot)$.

To optimize the efficiency of the offset compensator, sufficient consideration is given to the disparity existing between the initial sentence embedding and the retrieved sentence embedding.
Here, the distance between the embeddings is measured by mean squared error, which is written as
\begin{equation}\label{EM}
  \begin{aligned}
  \mathcal{L}^{\mathrm{SN}}(\{\mathbf{e}_{u}[i]\}, \{\widehat{\mathbf{e}}_{u}[i]\}; {\epsilon})= \frac{1}{\|\mathcal{U}\|}  \sum_u \Big\lVert \mathbf{e}_{u}[i]-\widehat{\mathbf{e}}_{u}[i] \Big\rVert_{2}^{2}.
  \end{aligned}
\end{equation}

Hence, the total loss of the whole semantic transmission network is denoted by
\begin{equation}\label{TOTAL}
  \begin{aligned}
  \mathcal{L}^{\mathrm{TOTAL}} = \underbrace{\mathcal{L}^{\mathrm{SE}}}_{\mathrm{reconstruction\ loss}} + \underbrace{\pounds \mathcal{L}^{\mathrm{SN}}}_{\mathrm{semantic\ noise\ loss}},
  \end{aligned}
\end{equation}
where $\pounds \in [0,1]$ is the weight coefficient.
\subsubsection{Gradient Patching}
In this paper, manual gradients are set based on trials with a small sample to repair the gradient, which is truncated by non-differentiable SBQ and OFDM transmission.
Despite the complexity that manual gradient setting brings to semantic network training, it can effectively solve the gradient vanishing problem.
Furthermore, while treating the gradient of wireless transmission as a constant, it is worth noting that the semantic information, along with the semantic noise arising from channel characteristics, e.g. error bits, can propagate within the semantic communication network and be captured by the compensator.

$\textbf{Algorithm}$ 2 demonstrates the details of training the offset compensator model, and
$\textbf{Algorithm}$ 3 show the details of tuning the whole semantic communication network. 
\begin{breakablealgorithm}
  \caption{Train Offset Compensator Model.}
  \label{alg:alg2}
  \begin{algorithmic}[1]
    \STATE \textbf{Input:} The knowledge vocabulary $V$, the resource allocation scheme $a_m$ and the parameter sets $\{{\beta},{\phi},{\eta},{\kappa}\}$.
    \STATE \textbf{Transmitter:} 
    \STATE \hspace{\algorithmicindent} Data source $V$ $\rightarrow$ $\{\mathbf{d}_u[i]\}$.
    \STATE \hspace{\algorithmicindent} $\mathcal{F}(\mathbf{d}_u[i];{\beta}) \rightarrow \{\mathbf{x}_u[i]\}$.
    \STATE \hspace{\algorithmicindent} $\mathcal{E}(\mathbf{x}_u[i];{\phi}) \rightarrow \{\mathbf{e}_u[i]\}$.
    \STATE \hspace{\algorithmicindent} Quantize $\{\mathbf{e}_u[i]\} \rightarrow \{\mathbf{e}_{u}^{\mathrm{bit}}[i]\}$.
    \STATE \textbf{Physical Channel:} 
    \STATE \hspace{\algorithmicindent} Transmit bit data $\{\mathbf{e}_{u}^{\mathrm{bit}}[i]\}$ with action $a_m$.
    \STATE \textbf{Receivers:} 
    \STATE \hspace{\algorithmicindent} Receive bit data $\{\widehat{\mathbf{e}}_u[i]\}$.
    \STATE \hspace{\algorithmicindent} Decode $\{\widehat{\mathbf{e}}_u[i]\} \rightarrow \{\mathring{\mathbf{e}}_u[i]\}$.
    \STATE \hspace{\algorithmicindent} $\mathcal{P}(\mathring{\mathbf{e}}_u[i];{\epsilon}) \rightarrow \{\widehat{\mathbf{e}}_u[i]\}$.
    \STATE \hspace{\algorithmicindent} Computer the loss $\mathcal{L}^{\mathrm{SN}}$ given by (\ref{EM}).
    \STATE \hspace{\algorithmicindent} Update ${\epsilon}$ with gradient descent.
    \STATE \textbf{Output:} The trained parameter set $\{{\epsilon}\}$.
  \end{algorithmic}
\end{breakablealgorithm}

\begin{breakablealgorithm}
  \caption{Tune Whole Semantic Representation Model.}
  \label{alg:alg3}
  \begin{algorithmic}[1]
    \STATE \textbf{Input:} The knowledge vocabulary $V$, the resource allocation scheme $a_m$ and the parameter sets $\{{\beta},{\phi},{\epsilon},{\eta},{\kappa}\}$.
    \STATE \textbf{Transmitter:} 
    \STATE \hspace{\algorithmicindent} Data source $V$ $\rightarrow$ $\{\mathbf{d}_u[i]\}$.
    \STATE \hspace{\algorithmicindent} $\mathcal{F}(\mathbf{d}_u[i];{\beta}) \rightarrow \{\mathbf{x}_u[i]\}$.
    \STATE \hspace{\algorithmicindent} $\mathcal{E}(\mathbf{x}_u[i];{\phi}) \rightarrow \{\mathbf{e}_u[i]\}$.
    \STATE \hspace{\algorithmicindent} Quantize $\{\mathbf{e}_u[i]\} \rightarrow \{\mathbf{e}_{u}^{\mathrm{bit}}[i]\}$.
    \STATE \textbf{Physical Channel:} 
    \STATE \hspace{\algorithmicindent} Transmit bit data $\{\mathbf{e}_{u}^{\mathrm{bit}}[i]\}$ with the action $a_m$.
    \STATE \textbf{Receivers:} 
    \STATE \hspace{\algorithmicindent} Receive bit data $\{\widehat{\mathbf{e}}_u[i]\}$.
    \STATE \hspace{\algorithmicindent} Decode $\{\widehat{\mathbf{e}}_u[i]\} \rightarrow \{\mathring{\mathbf{e}}_u[i]\}$.
    \STATE \hspace{\algorithmicindent} $\mathcal{P}(\mathring{\mathbf{e}}_u[i];{\epsilon}) \rightarrow \{\widehat{\mathbf{e}}_u[i]\}$.
    \STATE \hspace{\algorithmicindent} $\mathcal{E}^{-1}(\widehat{\mathbf{e}}_u[i];{\eta}) \rightarrow \{\widehat{\mathbf{x}}_u[i]\}$. 
    \STATE \hspace{\algorithmicindent} $\mathcal{F}^{-1}(\widehat{\mathbf{x}}_u[i];{\kappa}) \rightarrow \{\widehat{\mathbf{d}}_u[i]\}$. 
    \STATE \hspace{\algorithmicindent} Computer the loss $\mathcal{L}^{\mathrm{TOTAL}}$ given by (\ref{TOTAL}).
    \STATE \hspace{\algorithmicindent} Update ${\beta},{\phi},{\epsilon},{\eta},{\kappa}$ with gradient descent.
    \STATE \textbf{Output:} The trained parameter set $\{{\beta},{\phi},{\epsilon},{\eta},{\kappa}\}$ and $\mathcal{L}^{TOTAL}$.
  \end{algorithmic}
\end{breakablealgorithm}

\section{Performance Metrics and Problem Formulation}
\subsection{Performance Metrics}
This section introduces two essential metrics, namely, the semantic-bit mapping efficiency and semantic transmission delay,
provides a holistic evaluation of semantic communication systems, 
taking into account both the effectiveness of semantic information transmission and the timeliness of delivering the information to users.

\subsubsection{Semantic Quantization Efficiency}
The primary goal is to evaluate the similarity between the reconstructed and original data at the semantic level.
Inspired by \cite{9398576}, the enhanced ultra-large pre-trained model ALBERT \cite{lan2019albert} is introduced, 
where the ALBERT model can improve inference efficiency by significantly reducing the number of parameters in BERT, approximately by 18\%, while maintaining comparable performance levels.
The semantic similarity is written as
\begin{equation}
  \Im(\widehat{\mathbf{d}}_{u}[i],\mathbf{d}_{u}[i])=\frac{\boldsymbol{B}_{\boldsymbol{\varrho}}(\mathbf{d}_{u}[i]) \cdot \boldsymbol{B}_{\boldsymbol{\varrho}}(\widehat{\mathbf{d}}_{u}[i])^T}{\left\|\boldsymbol{B}_{\boldsymbol{\varrho}}(\mathbf{d}_{u}[i])\vphantom{\widehat{\mathbf{d}}_{u}[i]}\right\|\left\|\boldsymbol{B}_{\boldsymbol{\varrho}}(\widehat{\mathbf{d}}_{u}[i])\right\|},
\end{equation}
where $\boldsymbol{B}_{\boldsymbol{\varrho}}$ denotes the ALBERT model.
As pointed in \cite{9398576,lan2019albert,yang2019xlnet,liu2019roberta},
large-scale pre-trained models are trained on extensive corpora, enabling them to acquire rich language knowledge and semantic understanding. 
These models excel in capturing contextual information at the sentence level. 
In contrast to the traditional word matching-based approaches, these context-aware models exhibit a deeper understanding of sentence semantics and intent, 
resulting in a more accurate measurement of sentence similarity.

In order to address the trade-off between the semantic accuracy and the number of consumed bits, 
a new metric called semantic quantization efficiency (SQE) is proposed, which quantifies the ratio of the semantic similarity gain related to the bits for each semantic feature. 
For the $u$-th user during the transmission of the $i$-th text data, the SQE $\varpi_{u}[i]$ is represented by
\begin{equation}
  \varpi_u[i] = \frac{\Im(\widehat{\mathbf{d}}_{u}[i],\mathbf{d}_{u}[i])}{\sum_{n=1}^{N}\zeta_{u,n}n},
\end{equation}
where $\zeta_{u,n}$ denotes the allocation of the semantic-bit codec, $\zeta_{u,n}[i] \sim \pi_{\ddot{\boldsymbol{\theta}}}\left(\cdot \mid \mathbf{e}_{u}[i], \mathbf{h_{u}}[i] \right)$.
The $\mathbf{h_{u}}[i]$ denotes the channel state information at the $u$-th user during the transmission of the $i$-th data, and $\ddot{\boldsymbol{\theta}}$ denotes the parameter of the policy network. 
If the $u$-th user is paired with the $n$-th semantic-bit codec, $\zeta_{u,n}=1$; otherwise, $\zeta_{u,n}=0$. Each user is allocated with a semantic-bit codec.

\subsubsection{Semantic Transmission Delay}
In dynamic channel environments, the transmission delay of semantic information plays a crucial role in shaping the user experience, especially when different semantic encoding levels are assigned. 
To further optimize the transmission process, it is crucial to take the transmission delay as a key aspect of the QoS framework.
Furthermore, the inclusion of transmission delay enables the establishment of a real-time relationship between the channel state and semantic coding level assignment. 

The transmission rate of the $u$-th user over the subchannel $c$ during the transmission of the $i$-th text data is represented by
\begin{equation}
  \Re_{u,c}[i] = b_{u}[i] \log _2\left(1+\frac{\left|\mathbf{h}_{u, c}[i] \mathbf{f}_{u,c}[i]\right|^2}{\sigma_u^2[i]}\right),
\end{equation}
where $b_{u}$ denotes the bandwidth for the $u$-th user, and $\mathbf{h}_{u,c}^T \in \mathbb{C}^{1 \times M}$ and $\mathbf{f}_{u,c} \in \mathbb{C}^{M \times 1}$ 
respectively denote the channel and beamforming from the base station to the $u$-th user over the $c$-th subchannel.

Hence, the transmission latency at the $u$-th user during the transmission of the $i$-th raw data is represented by
\begin{equation}
  \mathcal{G}_u[i] =\frac{\sum_{n=1}^{N}\zeta_{u,n}n}{\sum_{c=1}^{C}\mathbf{\rho}_{u,c}[i]\Re_{u,c}[i]},
\end{equation}
where $\rho_{u,c}$ denotes the subchannel assignment. If the $u$-th user occupies the $c$-th subchannel, $\rho_{u,c}=1$; otherwise, $\rho_{u,c}=0$. 

\vspace{-0.25cm}
\subsection{Problem Formulation}
Based on the long-term text transmission, the probability of sentence occurrence in the specific knowledge base tends to stabilize.
Hence, the text dataset $\mathcal{I}$ can be sampled based on the sentence probability of each user.
The semantic latency at the $u$-th user is written as $\widehat{\mathcal{G}}_u = \frac{1}{\lvert\mathcal{I}\rvert} \sum_{i=1}^{I} \mathcal{G}_u[i].$

Considering that excessively low semantic similarity can give rise to potential misinterpretations during the semantic reconstruction,
the minimum tolerable similarity threshold $\Im^{th}$ is introduced, and the semantic transmission is valid only when the semantic similarity satisfies the minimum requirement $\Im^{th}$.
Hence, the effective SQE is formulated as $\widehat{\varpi}_u^{\sharp} = \frac{1}{\lvert\mathcal{I}\rvert} \sum_{i=1}^{I} \Im_u^{\sharp}[i] \varpi_u[i],$
where $\Im_u^{\sharp}[i]$ denotes the validity of semantic information during the transmission of the $i$-th data at the $u$-th user.
Specifically, if $\Im_u = \Im(\widehat{\mathbf{d}}_{u}[i],\mathbf{d}_{u}[i]) \geq \Im^{th}$, $\Im_u^{\sharp}[i]=1$; otherwise, $\Im_u^{\sharp}[i]=0$.

In the context of task-oriented semantic communication, attention should be given to manage semantic coding and resource allocation in a dynamic wireless communication environment. 
It is to find an optimal balance between maximizing the SQE and minimizing the latency, related to the semantic representation and resource allocation.
Hence, the SC-QoS is defined based on the SQE and transmission latency, and the effective SC-QoS is denoted by
\begin{equation}
  \Psi = \sum_{u=1}^{U} ( \widehat{\varpi}_u^{\sharp} - \phi_{\mathcal{G}} \widehat{\mathcal{G}}_u ),
\end{equation}
where $\phi_{\mathcal{G}}$ denotes the balance coefficient.

Let $\boldsymbol{\zeta}=\{\zeta_{u,n}\}_{\forall u \in U, \forall n \in N}$ denote the allocation vector of the number of the SBQ,
$\boldsymbol{\rho}=\{\rho_{u,c}\}_{\forall u \in U, \forall c \in C}$ denote the allocation vector of the subchannels,
$\boldsymbol{b}=\{b_{u}\}_{\forall u \in U}$ denote the allocation vector of the bandwidth resource,
and $\boldsymbol{F}=\{\mathbf{f}_{u,c}\}_{\forall u \in U, \forall c \in C}$ denote the transmission matrix from the base station. 
Based on the above system model, the optimization problem can be formulated as 
\begin{subequations}\label{op}
  \begin{align}
   \mathbf{P}: &\max_{\boldsymbol{\zeta}, \boldsymbol{\rho}, \boldsymbol{b} ,\boldsymbol{F}} \Psi \\ 
   \text{ s.t. }  
   &\zeta_{u, n} \in\{0,1\}, \forall u \in \mathcal{U}, \forall n \in \mathcal{N},\\
   &\sum_{n=1}^N\zeta_{u, n}=1, \forall u \in \mathcal{U},\\
   &\rho_{u, c} \in\{0,1\}, \forall c \in \mathcal{C}, \forall u \in \mathcal{U},\\
   &\sum_{c=1}^C\rho_{u, c}=1, \forall u \in \mathcal{U},\\
   &\sum_{u=1}^U b_{u} \leq B,\\
   &\sum_{u=1}^U\sum_{c=1}^C\left\|\mathbf{f}_{u,c}\right\|^2 \leq TP,\\
   &\widehat{\mathcal{G}}_u \leq \mathcal{G}^{\mathrm{th}}, \forall u \in \mathcal{U},\\
   &\widehat{\Im}_u \geq \Im^{\mathrm{th}}, \forall u \in \mathcal{U},
  \end{align}
\end{subequations}
where $\mathcal{G}_u^{\mathrm{th}}$ represents the maximum tolerance for latency, and $\Im^{\mathrm{th}}$ is the minimum requirements of the semantic similarity.
Constraint (\ref*{op}b) and (\ref*{op}c) represent the constraints related to the allocation of semantic quantization. 
Constraint (\ref{op}d) and Constraint (\ref{op}e) represent the constraints regarding the allocation of subchannels to users.
(\ref{op}f) is the constraint of the total bandwidth.
(\ref{op}g) represents the constraint on the total transmit power of the base station.
Constraint (\ref*{op}h) imposes a restriction on the maximum transmission latency considering the QoS of the users.
Constraint (\ref*{op}i) imposes restrictions on the semantic similarity.

\section{Proposed Intelligent Resource Allocation Scheme}
\begin{figure*}
  \centering
  \includegraphics[scale=0.4]{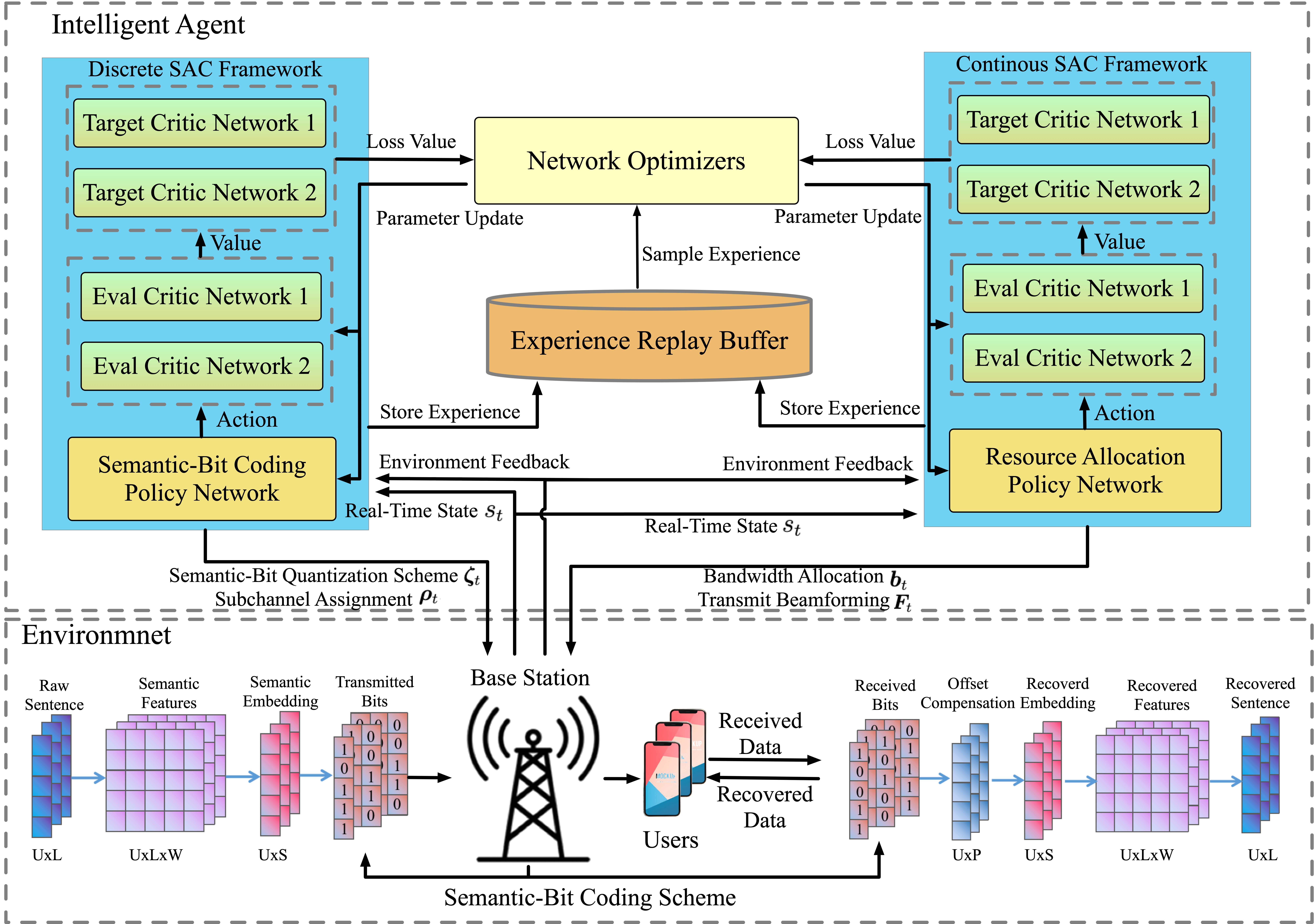}
  \caption{The proposed adaptive intelligent resource allocation scheme for wireless semantic communications.}
  \vspace{-0.4cm}
\end{figure*}
In this section, we propose an adapative environment-aware intelligent resource allocation scheme without mapping guidance for semantic communication, as shown in Fig. 3.

\subsection{MDP Problem Modeling}
The Markov decision process (MDP) serves as a formal approach that offers a mathematical framework and modeling tools derived from the reinforcement learning theory to describe sequential decision problems. 
To ensure the efficiency of the RL algorithm, the optimization problem in (\ref{op}) is formulated as an MDP problem. 
In our proposed model, the semantic communication system is treated as the environment, and the intelligent control unit of the base station is considered as the intelligent agent. 
The following elaborates on the other essential components of our proposed RL problem, including the state space, action space, reward function, and transition probabilities.

\textbf{State space:} The state space of our proposed semantic communication network is denoted by $\mathcal{S}$, and the system state is written as $s_t$, $s_t \in \mathcal{S}$. 
Let $\mathbf{h}_{u,t}$ denote the exact channel coefficients at the $u$-th user.
To perceive the semantic task and take advantage of the channel characteristics, the normalized sentence embedding information and channel coefficients are incorporated into the system state, which are respectively represented by
\begin{subequations}
  \begin{align}
  \mathbf{e}_t \triangleq \left\{\frac{\mathbf{e}_{1,t}}{\left\|\mathbf{e}_{1,t}\right\|},\frac{\mathbf{e}_{2,t}}{\left\|\mathbf{e}_{2,t}\right\|},\cdots,\frac{\mathbf{e}_{u,t}}{\left\|\mathbf{e}_{u,t}\right\|},\cdots, \frac{\mathbf{e}_{U,t}}{\left\|\mathbf{e}_{U,t}\right\|}\right\},\\
  \mathbf{h}_t \triangleq \left\{\frac{\mathbf{h}_{1,t}}{\left\|\mathbf{h}_{1,t}\right\|},\frac{\mathbf{h}_{2,t}}{\left\|\mathbf{h}_{2,t}\right\|},\cdots,\frac{\mathbf{h}_{u,t}}{\left\|\mathbf{h}_{u,t}\right\|},\cdots, \frac{\mathbf{h}_{U,t}}{\left\|\mathbf{h}_{U,t}\right\|}\right\}.
  \end{align}
\end{subequations}
The other key elements in the state $s_t$ are the selected actions $a_{t-1}$, semantic features efficiency $\boldsymbol{\varpi}_t = \{\varpi_{u,t}\}_{\forall u \in \mathcal{U}}$, transmission latency $\boldsymbol{\mathcal{G}_t} =\{\mathcal{G}_{u,t}\}_{\forall u \in \mathcal{U}}$, and the reward $r_{t}$ obtained up to the current time step. 
Therefore, the system state $s_t$ is written as $s_t =\biggl\{a_{t-1},\mathbf{e}_{t}, \mathbf{h}_{t}, \boldsymbol{\varpi}_{t}, \boldsymbol{\mathcal{G}}_{t}, r_{t}\biggr\}.$

\textbf{Action space:} The action space of the network is denoted by $\mathcal{A}$, and the action selected at the timestamp $t$ is written as $a_t$, $a_t \in \mathcal{A}$.
The action $a_t$ encompasses the SBQ scheme $\boldsymbol{\zeta}_t$, the subchannel assignment $\boldsymbol{\rho}_t$, the bandwidth allocation $\boldsymbol{b}_t$, and the transmit beamforming $\boldsymbol{F}_t$,
which is represented by $a_t=\biggl\{\boldsymbol{\zeta}_t, \boldsymbol{\rho}_t, \boldsymbol{b}_t, \boldsymbol{F}_t \biggr\}.$

\textbf{Transition probability:} The state transition probability in reinforcement learning is represented by $\mathcal{V}\left(s_{t+1} | s_t, a_t\right)$, which indicates the probability of transitioning from state $s_t$ to state $s_{t+1}$ given action $a_t$,
associated with the occurrence probability of the current transmitted sentence in the knowledge base for each user.

\textbf{Reward function:} The key to the design of the reward function is to guide the agent to learn in the direction of maximizing SC-QoS, 
providing immediate feedback signals to the agent. 
Based on the objectives and constraints, an efficient reward function, 
incorporating the requirements for the SFE, transmission latency and semantic similarity, is designed to constrain the behavior of the agent and mitigate the reward sparsity problem.
Hence, the reward function is designed as $r = \Psi + \omega_{\rm{P}}^{\Im} \sum_{u=1}^{U} \rm{P}_u^{\Im} + \omega_{\rm{P}}^{\mathcal{G}} \sum_{u=1}^{U}\rm{P}_u^{\mathcal{G}},$
where $\rm{P}_u^{\varpi}$ and $\rm{P}_u^{\mathcal{G}}$ are the punishments for not fulfilling the requirements for the semantic similarity and transmission latency at the $u$-th user, respectively.
The $\omega_{\rm{P}}^{\varpi}$ and $\omega_{\rm{P}}^{\mathcal{G}}$ are the given coefficients of the punishments.
For better evaluation, the punishments are designed by
\begin{subequations}
  \begin{align}
  &\rm{P}_u^{\Im} = \begin{cases} \mathring{r}, & \Im_u \geq \Im^{th}, \\ \Im^{th} - \Im_u, & 0 \leq \Im_u < \varpi^{th},\end{cases}\\
  &\rm{P}_u^{\mathcal{G}} = \begin{cases} \mathring{r}, & \mathcal{G}_u \geq \mathcal{G}^{th}, \\ \mathcal{G}^{th} - \mathcal{G}_u, & 0 \leq \mathcal{G}_u < \mathcal{G}^{th},\end{cases}
\end{align}
\end{subequations}
where $\mathring{r}$ is a positive reward with given small value. 

\vspace{-0.2cm}
\subsection{Bellman Equation And Policy Gradient}
The Bellman equation offers an approximation method that relies on a value function, providing accurate feedback to the base station. 
To find an optimal strategy that maximizes the long-term performance of the semantic communication network, the cumulative discount reward at time step $t$ is written as
$R(s,a)=\sum_{\varepsilon=0}^{\infty} \gamma^\varepsilon r_{t+\varepsilon+1}(s,a)$,
where $\gamma \in[0,1)$ is the discount rate, which determines the relative weight of future rewards and addresses the issue of delayed rewards.
The state-action value function $Q_\pi(s, a)$ with the policy $\pi$ is written as $Q_\pi\left(s, a\right)=\mathbb{E}_{a \sim \pi}\left[R(s,a)\right]$.
Then, the $Q$ function based on the Bellmann equation is represented by
$
Q_\pi\left(s, a\right)=\mathbb{E}_{a \sim \pi}\left[r(s, a)\right]+\gamma \sum_{s^{\prime} \in S} \mathcal{V}\left(s^{\prime} \mid s, a\right)\left[\sum_{a^{\prime} \in A} \pi\left(s^{\prime}, a^{\prime}\right) Q_\pi\left(s^{\prime}, a^{\prime}\right)\right].
$

The policy gradient method is capable of adapting to nondeterministic environments by learning the probability distribution of strategies within high-dimensional state and action spaces. 
This makes it suitable for handling problems with high-dimensional and continuous action spaces formulated in this paper.
Mathematically, this can be expressed as
$\vartheta^*=\arg \max _\vartheta \mathbb{E}_{\tau \sim \pi_\vartheta}\left[\sum_t^{\infty} \gamma^t r_t\right]$,
where $\tau$ is the decision trajectory of the base station.

In this study, the Bellman equation and policy gradient techniques are jointly employed to precisely evaluate the performance of the semantic communication network, 
thereby facilitating an optimal resource allocation scheme for semantic communication networks.

\vspace{-0.2cm}
\subsection{SAC Algorithm for Bandwidth Allocation and Transmit Beamforming}
The soft actor-critic (SAC) algorithm \cite{sac2} combines the principles of the Bellman equation and the policy gradient approach, effectively balancing exploration and exploitation in resource allocation for semantic tasks.
Furthermore, the maximum entropy is introduced in SAC which enhances the exploration capability and prevents the agent from trapping into the local optima. 
Hence, the SAC is employed to optimize the bandwidth allocation and the transmit beamforming.

The entropy regularization of the optimal strategy parameter is expressed as
$
  \vartheta^*=\underset{\vartheta}{\operatorname{argmax}} \underset{\substack{\tau \sim \pi_\vartheta \\ (s_t,a_t) \sim \tau}} {\mathbb{E}}\left[\sum_{t=0}^{\infty} \gamma^t\left(r_t(s_t,a_t) + \alpha H\left(\pi_\vartheta\left(\cdot \mid s_t\right)\right)\right)\right],
$
where $\pi_\vartheta\left(\cdot \mid s_t\right)$ denotes the probability density function of the action distribution at the state $s_t$ with the parameter $\vartheta$, 
and $\alpha$ stands for temperature, which determines the learnable trade-off between the obtained reward and the policy randomness.
Similarly, the entropy term is added to the Bellman equation.
The entropy regularized Bellman equation for the optimal Q-function is written as
$
  Q(s_t, a_t)=\underset{\substack{a_{t+1} \sim \pi \\ s_{t+1}\sim P_{a_{t+1}}}}{\mathbb{E}}\left[r\left(s_t, a_t\right)+\gamma\left(Q\left(s_{t+1}, a_{t+1}\right)+\alpha H\left(\pi\left(\cdot \mid s_t\right)\right)\right)\right],
$
where $P_{a_{t+1}}$ denotes the distribution of reachable states when the $a_{t+1}$ is operated.

As SAC is built upon a stochastic policy, sampling an action from a traditional distribution defined by policy parameters $\theta$ and 
state $s$ results in lost gradients, which is represented by $\check{a}^{\prime} \sim \mathcal{N}\left(\mu_{\sigma_{\pi_\vartheta}}(s), \sigma_{\pi_\vartheta}(s)\right)$. 
Given the mean $\mu(\sigma_{\pi_\vartheta}\left(s\right))$ and variance $\sigma_{\pi_\vartheta\left(s\right)}$ output from the policy network, 
the reparameterization technique is employed in SAC to promote exploration and establish gradient connectivity, which is expressed as
$
\check{a}_{\pi_\vartheta}(s, \xi)=\tanh \left(\mu_{\pi_\vartheta}(s)+\sigma_{\pi_\vartheta}(s) \odot \xi\right),
$
where $\xi \sim \mathcal{N}(0,1)$.

A sample $D$ with batch size $Z$ is randomly drawn from the experience replay pool,
which is represented by $D=\{\left(s_z, a_z, r_{z+1}, s_{z+1}\right)\}, z \in Z$.
Based on entropy regularization, the loss function of the actor network is represented by
\begin{equation}\label{loss1}
  \mathcal{J}_{\S}(\theta)=\underset{\substack{s_z \sim D \\ \xi \sim \mathcal{N}}}{\mathbb{E}}\left[-\underset{j=1,2}{\min}Q_{\S_j}\left(s_z, \check{a}_\theta(s_z, \xi)\right)+\alpha \log \pi_\theta\left(\check{a}_\theta(s_z, \xi) \right)\right],
\end{equation}
where $\theta$ represents the parameter of the actor network, and $\S_j$ represents the parameter of the $j$-th critic network.
The loss function of the critic networks with entropy regularization is expressed as
\begin{equation}\label{loss2}
  \begin{aligned}
  &\mathcal{J}_{\pi_{\theta}}(\S)=\underset{\substack{\left(s_z, a_z, r_z, s_{z+1}, d\right) \sim D\\\check{a}_{z+1} \sim \pi_\theta\left(\cdot \mid s_{z+1}\right)}}{\mathbb{E}}[(Q_{\S_j}(s, a)-(r+\\
  &\gamma(\underset{j=1,2}{\min}Q_{\S_j}(s_{z}, a_{z})-\alpha \log \pi_\theta(\check{a}_{z+1}\mid s_{z+1}))))^2].
  \end{aligned}
\end{equation}
To learn an optimal temperature parameters, the loss function is written as
\begin{equation}\label{loss3}
  \mathcal{J}_{\theta,\S}(\alpha)=\underset{\substack{s_z \sim D \\ \check{a}_z \sim \pi_\theta}}{\mathbb{E}}\left[-\alpha \log \pi_\theta(\check{a}_z \mid s_z)+\tilde{H}\right],
\end{equation}
where the term $\tilde{H}$ represents the target entropy and is the given hyperparameter.

\vspace{-0.35cm}
\subsection{D-SAC Algorithm for SBQ and Subchannel Assignment}
Building upon the mentioned advantages of the SAC algorithm, it is extended to the discrete space to tackle the SBQ and the subchannel assignment.

The loss function of the actor network for discrete actions is expressed as
\begin{equation}\label{loss4}
  \mathcal{J}_{Q_{\ddot{\S}_j}}(\ddot{\theta})=\underset{s_z \sim D}{\mathbb{E}}\left[\pi_{\ddot{\boldsymbol{\theta}}}(s_z)^T\left(-\underset{j=1,2}{\min}Q_{\ddot{\S}_j}(s_z)+\ddot{\alpha} \log \pi_{\ddot{\theta}}(s_z)\right)\right],
\end{equation}
where $\ddot{\theta}$ represents the parameter of the actor network, $\ddot{\S}_j$ represents the parameter of the $j$-th critic network in D-SAC and $\ddot{\alpha}$ represents the learnable temperature.
Two critic network structures are designed to mitigate the Q-value overestimation problem.
The loss function of the critic networks for discrete actions is represented by
\begin{equation}\label{loss5}
  \begin{aligned}
  &\mathcal{J}_{\pi_{\ddot{\theta}}}(\ddot{\S}_j)=\underset{\left(s_z, a_z, r_z, s_{z+1}\right) \sim D}{\mathbb{E}}[(Q_{\ddot{\S}_j}(s_z)-(r_z\\
  &+\gamma(\pi_{\ddot{\theta}}(s_z)(Q_{\ddot{\S}_j}(s_{z+1})-\ddot{\alpha} \log \pi_{\ddot{\theta}}(s_{z+1})))))^2].
  \end{aligned}
\end{equation}
The loss function of the temperature for discrete actions is written as
\begin{equation}\label{loss6}
  \mathcal{J}_{\ddot{\alpha}}(\ddot{\alpha})=\underset{s_z \sim D}{\mathbb{E}}\left[\pi_{\ddot{\theta}}(s_z)^T\left(-\ddot{\alpha} \log \pi_{\ddot{\boldsymbol{\theta}}}(s_z)+\hat{H}\right)\right],
\end{equation}
where the term $\hat{H}$ represents the target entropy and is the given hyperparameter in D-SAC.

The details of our proposed scheme are summarized in $\textbf{Algorithm}$ 4.
\begin{breakablealgorithm}
  \caption{Our Designed Adaptive Semantic Resource Allocation.}
  \label{alg:alg0}
  \begin{algorithmic}[1]
    \STATE \textbf{Initialize:} The semantic coding parameters ${\beta},{\phi},{\epsilon},{\eta},{\kappa}$, 
    the semantic resource allocation parameters $\theta,\S_1,\S_2,\alpha,\ddot{\theta},\ddot{\S}_1,\ddot{\S}_2,\ddot{\alpha}$,
    the experience replay buffer $D$, episode $T$, step per episode $M$, batch size $Z$ for training and the knowledge vocabulary $V$.
    \STATE \textbf{Copy:} The target parameters $\theta^{-} \leftarrow \theta, \S_1^{-} \leftarrow \S_1, \S_2^{-} \leftarrow \S_2, \ddot{\theta}^{-} \leftarrow \ddot{\theta}, \ddot{\S}_1^{-} \leftarrow \ddot{\S}_1, \ddot{\S}_2^{-} \leftarrow \ddot{\S}_2$.
    \STATE Train the semantic coding model, given in $\textbf{Algorithm}$ 1.
  \FOR{episode $t=1 \rightarrow T$}
          \STATE Intial the action with random strategy: $a_0=\left\{\boldsymbol{\zeta}_0, \boldsymbol{\rho}_0, \boldsymbol{b}_0, \boldsymbol{F}_0\right\}$.
          \STATE Intial the state based on the intial action: $s_0=\biggl\{\{\mathbf{e}_{u,0}\}, \{\mathbf{h}_{u,0}\}, \{\varpi_{u,0}\}, \{\mathcal{G}_{u,0}\}\biggr\}$.
          \FOR{Step $m=1$ $\rightarrow$ $M$}
          \STATE Obtain discrete actions $a_m^d=\left\{\boldsymbol{\rho}_m,\boldsymbol{\zeta}_m\right\}$ and continuous actions $a_m^c=\left\{\boldsymbol{b}_m, \boldsymbol{F}_m\right\}$.
          \STATE Train the offset compensator model with $a_m=\{a_m^c,a_m^d\}$, presented in $\textbf{Algorithm}$ 2.
          \STATE Tune the whole semantic communication modularity with $a_m=\{a_m^c,a_m^d\}$, presented in $\textbf{Algorithm}$ 3.
          \STATE Execute $a_m$, the state turns into $s_{m+1}=\biggl\{\{\mathbf{x}_{u,m+1}\}, \{\mathbf{h}_{u,m+1}\}, \{\varpi_{u,m+1}\}, \{\mathcal{G}_{u,m+1}\}\biggr\}$.
          \STATE Obtain the reward $r_m$ and Store $\left(s_m, a_m, r_m, s_{m+1}\right)$ in the experience replay buffer $D$.
          \IF {the size of the experience replay buffer $\geq Z$}
          \STATE Sample from the $D$: $\left\{\left\{s_z, a_z, r_z, s_{z+1}\right\}\right\}_{z=1, \ldots, Z}$.
          \STATE Update $\theta$, $\S$ and $\alpha$ with the gradient $\nabla \mathcal{J}_{Q_{\S_j}}(\theta)$, $\nabla \mathcal{J}_{\pi_\theta}(\S)$ and $\nabla \mathcal{J}_{\alpha}(\alpha)$.
          \STATE Update $\ddot{\theta}$, $\ddot{\S}$ and $\ddot{\alpha}$ with the gradient $\nabla \mathcal{J}_{Q_{\ddot{\S}_j}}(\ddot{\theta})$, $\nabla (\ddot{\alpha})\mathcal{J}_{\pi_{\ddot{\boldsymbol{\theta}}}}(\ddot{\S})$ and $\nabla \mathcal{J}_{\ddot{\alpha}}(\ddot{\alpha})$.
          \STATE Updated target parameters every $K$ steps: $\theta^{-} \leftarrow \theta$, $\S \leftarrow \S$, $\ddot{\theta}^{-} \leftarrow \ddot{\theta}$ and $\ddot{\S}^{-} \leftarrow \ddot{\S}$.
          \ENDIF
      \ENDFOR
  \ENDFOR.
  \end{algorithmic}
\end{breakablealgorithm}

\section{Simulation Results}
This section evaluates the performance of our proposed adaptive intelligent resource allocation scheme for semantic communication networks.
The simulation parameters of the network are established as follows unless otherwise noted, which are set based on the settings used in \cite{9398576} and \cite{wang2023intelligent}.
The parameter setting is given in $\textbf{Table I}$.
The channels from the base station to users are assumed to be Rician fading,  
and the path fading is set as $PL = \left(PL^{\mathrm{ref}} - 10 \varphi \log _{10}\left(d / D^{\mathrm{ref}}\right)\right) \mathrm{~dB}$.
The structure of our designed semantic communication network is illustrated in $\textbf{Table II}$.
\begin{table*}
  \caption{Parameter Setting}
  \vspace{-0.3cm}
  \centering
  \begin{adjustbox}{width=450px}
  \begin{tabular}{cccc}
    \multicolumn{4}{c}{}{}\\
    \multicolumn{4}{c}{\textbf{System Parameters}}\\
    \hline
    Simulation Parameter& Value& Simulation Parameter& Value\\
    \hline
    The number of the users, $U$& 3 & The number of the subchannels, $C$& 3\\
    \hline
    The number of the base station antennas, $M$& 6 &The transmit power of the base station, ${PT}_b$& $-10\mathrm{~dBm}$\\
    \hline
    The total channel bandwidth, $B$& $90\mathrm{~kHz}$ & The path loss exponent from the base station to users, $\varphi$& 3\\
    \hline
    The variance of the noise at the users, $\sigma_u^2$ & 0.01 & The reference for path loss, $PL^{\mathrm{ref}}$& $30 \mathrm{~dB}$ \\
    \hline
    The reference for distance, $D^{\mathrm{ref}}$& $1 \mathrm{~m}$ & The modulation & 16-QAM\\
    \hline
    \multicolumn{4}{c}{}{}\\
    \multicolumn{4}{c}{\textbf{Semantic Network Parameters}}\\
    \hline
    Simulation Parameter& Value& Simulation Parameter& Value\\
    \hline
    The number of SBQ schemes, $N$& 3 &  The weight coefficient, $\pounds$ & $10^{-1}$ \\
    \hline
    The optimizer & Adam & The learning rate& $10^{-4}$  \\
    \hline
    The weight decay & $5 \times 10^{-4}$ & The decay factor & (0.9, 0.97)\\
    \hline
    \multicolumn{4}{c}{}{}\\
    \multicolumn{4}{c}{\textbf{DRL Network Parameters}}\\
    \hline
    Simulation Parameter& Value& Simulation Parameter& Value\\
    \hline
    The reward discount, $\gamma$ & 0.99 & The soft update parameter & $5 \times 10^{-3}$ \\
    \hline
    The target entropy, $\hat{H}$ and $\tilde{H}$  & -1 &The size of the memory replay & $2\times10^4$ \\
    \hline
    The learning rate of actor networks& $10^{-4}$ & The learning rate of critic networks& $10^{-4}$\\
    \hline
    The learning rate of entropy networks& $10^{-4}$ & The number of the hidden layers for each network & 3\\
    \hline
    The number of the neurons for each layers & 512 & The semantic similarity punishment coefficient, $\omega_{\rm{P}}^{\Im}$& 1 \\
    \hline
    The SQE punishment coefficient, $\omega_{\rm{P}}^{\mathcal{G}}$& 10 & The reward in punishment, $\mathring{r}$& 3\\
    \hline
  \end{tabular}
\end{adjustbox}
\end{table*}

\begin{table*}
  \caption{The Structure Of Our Proposed Separate Semantic Communication Network}
  \centering
  \begin{tabular}{|c|c|c|c|}
    \hline
    {}& Layer &Output Size &Activation\\
    \hline
    Semantic Encoder&Transformer Encoder&30$\times$128&Linear\\
    \hline
    \multirow{3}*{Sentence Semantic Embedder}&Dense &30$\times$512 &Relu\\
    \cline{2-4}
    &Dense &30$\times$128 &Relu\\
    \cline{2-4}
    &Pool&1$\times$64  &Linear\\
    \hline
    \multirow{2}*{Semantic-Bit Codec}&Semantic-Bit Encoder &None &None\\
    \cline{2-4}
    &Semantic-Bit Decoder &None &None\\
    \hline
    \multirow{2}*{Offset Compensator}&Dense &1$\times$512 & Relu\\
    \cline{2-4}
    &Dense &1$\times$64 &Linear\\
    \hline
    \multirow{4}*{Semantic Inverse Embedder}&Unpool&30$\times$128  &Linear\\
    \cline{2-4}
    &Dense &30$\times$512 &Relu\\
    \cline{2-4}
    &Dense &30$\times$128 &Linear\\
    \cline{2-4}
    &Residule Function &30$\times$128 & Normalization\\
    \hline
    \multirow{2}*{Semantic Decoder}&Transformer Decoder&30$\times$128&Linear\\
    \cline{2-4}
    &Predicter& Knowledge Sentence Size&Softmax\\
    \hline
    \end{tabular}
    \vspace{-0.15cm}
\end{table*}

\vspace{-0.3cm}
\subsection{Algorithm Training and Performance Comparison}
\begin{figure}
  \centering
  \includegraphics[width=9cm]{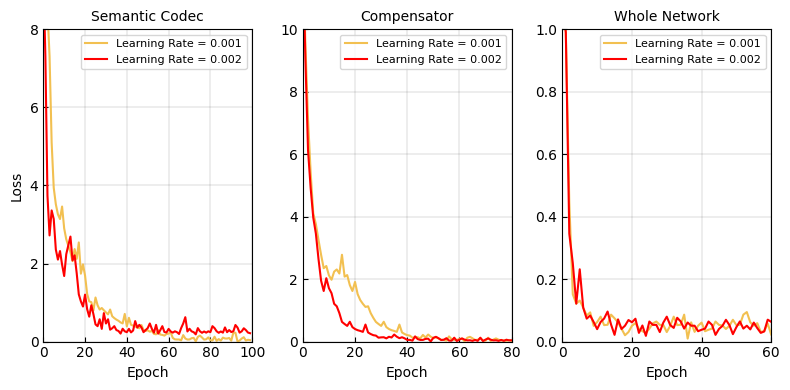}
  \vspace{-0.4cm}
  \caption{The impact of different learning rates during different training phases, $\Im^{\mathrm{th}}=0.8$.}
  \vspace{-0.4cm}
\end{figure}
Fig. 4 clearly illustrates the loss value versus the epochs during the training of the semantic representation, the compensator and the whole network, respectively.
A learning rate of 0.002 is chosen to train the semantic compensator due to its superior convergence performance as compared to that of 0.001.
Conversely, the learning rate of 0.001 is employed to train the semantic codec and fine-tune the entire network, aiming to improve the training performance. 

\begin{figure}
  \centering
  \setlength{\subfigcapskip}{-0.3cm}
  \subfigure[The reward of different users versus iterations.]{
    \begin{minipage}{8cm}
    \includegraphics[width=\textwidth]{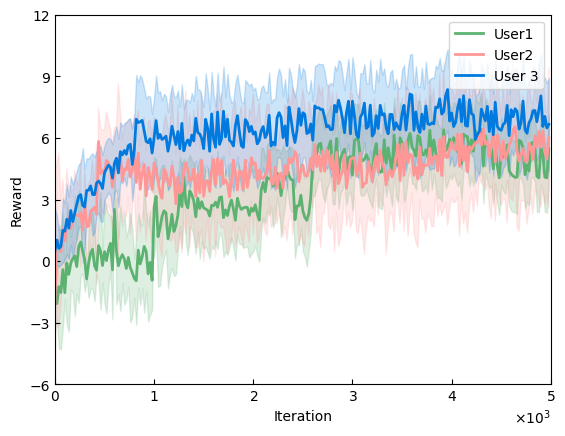} \\
    \end{minipage}}
  \vspace{-0.1cm} 
  \subfigure[The reward of different algorithms versus iterations.]{
    \begin{minipage}{8cm}
      \includegraphics[width=\textwidth]{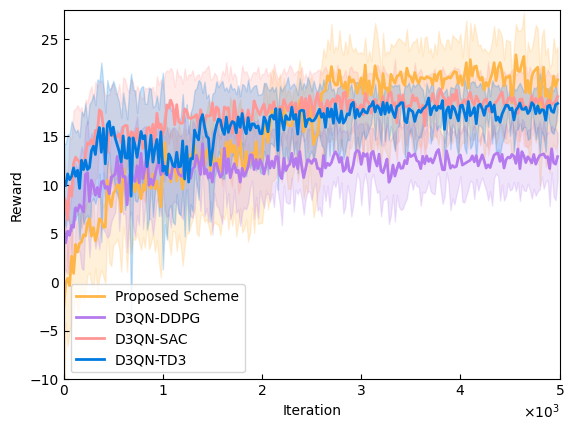} \\
    \end{minipage}}
  \caption{The local and global convergence of our proposed intelligent schemes.}
  \vspace{-0.2cm} 
\end{figure}

\begin{table}
  \caption{The Introduction Of The Compared Algorithms}
  \centering
  \begin{tabular}{|c|c|c|c|}
    \hline
    Algorithm& Network &Learning Rate &Space Scope\\
    \hline
    D3QN& Dueling-Deep Q Network &$3 \times 10^{-4}$ & Discrete\\
    \hline
    \multirow{2}*{DDPG}& Policy Network&$2 \times 10^{-4}$ & \multirow{2}*{Continuous}\\
    \cline{2-3}
    &Q-Value Network&$2 \times 10^{-4}$ & \\
    \hline 
    \multirow{2}*{TD3}& Actor Network  &$3 \times 10^{-4}$ & \multirow{2}*{Continuous}\\
    \cline{2-3}
    & Critic Q Networks & $3 \times 10^{-4}$ & \\
    \hline
  \end{tabular}
  \vspace{-0.3cm}
\end{table}
Fig. 5 demonstrates the convergence of our proposed environment-aware intelligent resource allocation scheme for semantic communication networks.
The dueling double deep Q-network (D3QN) algorithm, the DDPG algorithm and the twin delayed deep deterministic policy gradient (TD3) algorithm are considered for comparison.
The number of the network layers and the hidden neurons in D3QN, DDPG and TD3 is the same as the setting of SAC. 
The details are shown in $\textbf{Table III}$.
From Fig. 5(a), it can be seen that our proposed algorithm can achieve dynamic resource managements in accordance with the collective interests of the semantic communication network.
This is because our designed scheme can effectively explore the high-dimensional action and state space, and achieve the optimality of the formulated problem.
From Fig. 5(b), compared to the benchmark algorithms, our proposed scheme using the SAC and D-SAC achieves better SC-QoS. 
This is due to the fact that the maximum entropy-based SAC algorithm has a better exploration ability, 
which renders our proposed scheme suitable for intelligent applications in semantic communication networks.
While certain alternative base schemes can achieve convergence within a relatively short epoch, their effectiveness in discovering suboptimal solutions for resource allocation remains limited.
This deficiency is due to the insufficient exploration capabilities, and therefore alternative schemes are not applied in this paper.

\vspace{-0.1cm}
\subsection{SBQ and Semantic Noise Adversarial}
\begin{figure}
  \centering
  \includegraphics[width=7.8cm]{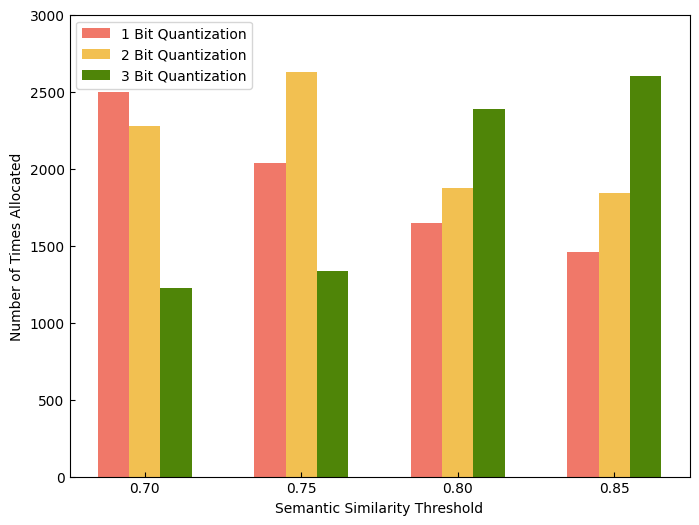}
  \vspace{-0.1cm}
  \caption{The SBQ with our proposed allocation scheme.}
\end{figure}

\begin{figure}
  \centering
  \includegraphics[width=7.8cm]{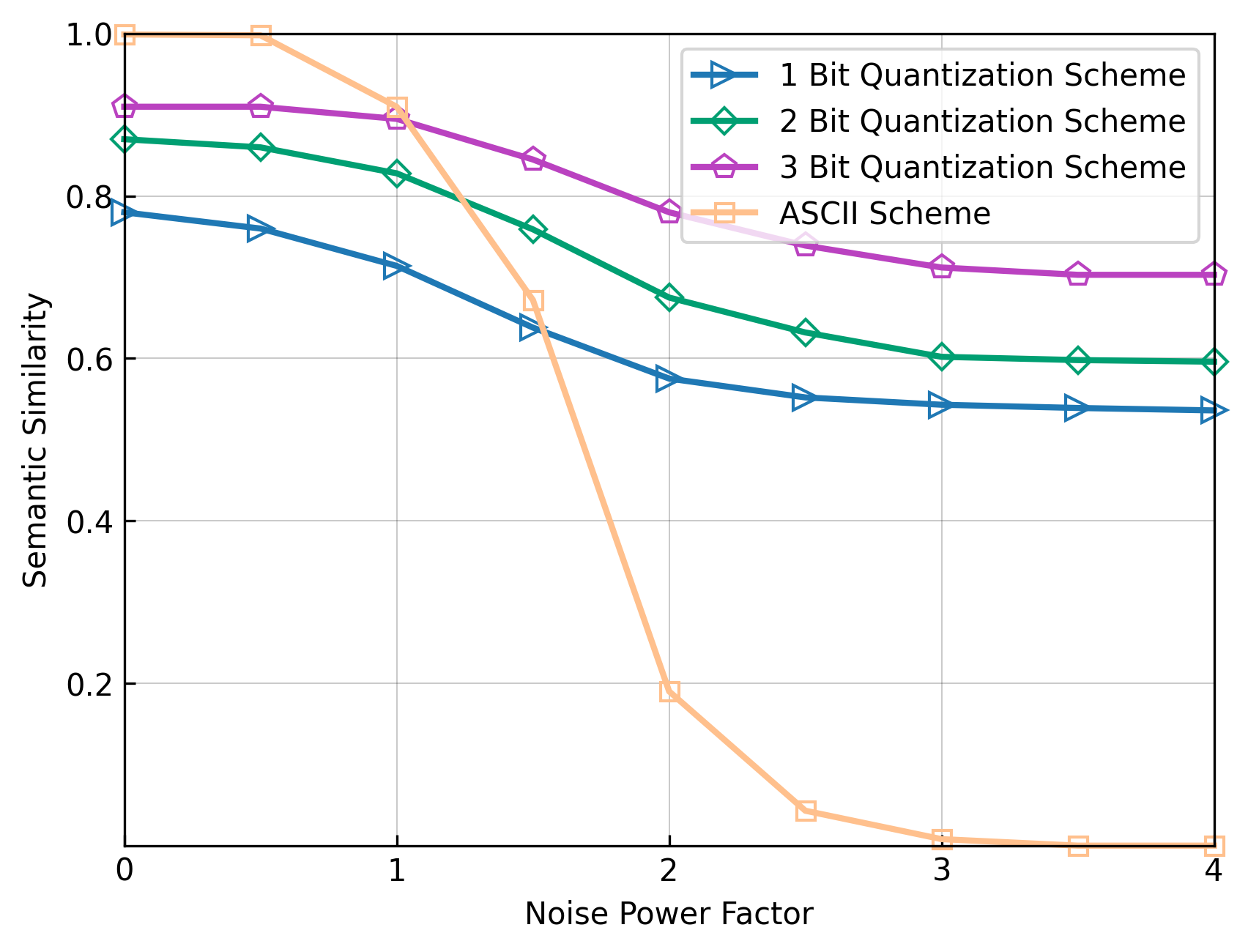}
  \caption{The semantic similarity versus the noise power.}
  \vspace{-0.15cm}
\end{figure}

\begin{table*}
  \caption{The target sentences and reconstructed sentence with different quantization schemes}
  \centering
  \begin{tabular}{|m{5cm}|m{5cm}|m{4cm}|}
    \hline
    Target Sentence& Reconstructed Sentence & Semantic Quantization Scheme\\
    \hline
    i will agree with the resolution & i will agree with the resolution & 1 Bit Quantization\\
    \hline
    there is no lack of seriousness and toughness who can think otherwise &  there was no lack of seriousness and toughness who could think otherwise & 2 Bit Quantization\\
    \hline
    you now have the chance to reply to this block of ten questions and to these seven questioners & you now have the chance to reply to the block of ten questions and to these seven questioners  & 3 Bit Quantization\\
    \hline
\end{tabular}
\vspace{-0.25cm}
\end{table*}
Fig. 6 illustrates the dynamic allocation of SBQ schemes across the semantic knowledge base, 
providing insights into the variation and distribution of SBQ with our proposed resource allocation scheme.
From Fig. 6, the SBQ can adaptively utilize more bits as the semantic similarity threshold increases to enhance the SC-QoS. 
Additionally, the usage of 2-bit quantization shows an upward trend within the range of semantic similarity thresholds from 0.70 to 0.75 but experiences a decline beyond that threshold range. 
This observation indicates that the preference for different quantization levels is influenced by the threshold, related to the specific requirements of semantic communications.
An example of suboptimal SBQ is presented in $\textbf{Table IV}$, where the quantization scheme depends on various factors, including the semantic characteristics (such as sentence length and semantic information richness) and the wireless communication environment. 

Fig. 7 depicts the semantic similarity versus the noise power, where the ASCII encodes each word into bits.
The noise power factor $p_f$ is set to adjust the noise power ${PT}_n$, denoted by ${PT}_n={10}^{13+p_f}PT_0$, where $PT_0 = 10^{(N_0 - 30) / 10}$.
From Fig. 7, it can be seen that the semantic similarity decreases with the escalation of noise power, and more accurate reconstructions can be obtained by using more bits with our proposed SBQ.
Notably, even under conditions of substantial noise, our proposed semantic communication network exhibits robust anti-interference properties.
This is due to the fact that the inherent noise in the forward pass of neural networks provides a basis for effectively mitigating the semantic accuracy loss problem. 
Moreover, for conventional coding represented by ASCII, a significant loss of semantic similarity occurs as the noise power increases, primarily due to the demand for the symbol-level accuracy and limited error correction.

\subsection{SC-QoS with Limited Communication Resources}
\begin{figure}
  \centering
  \includegraphics[width=8cm]{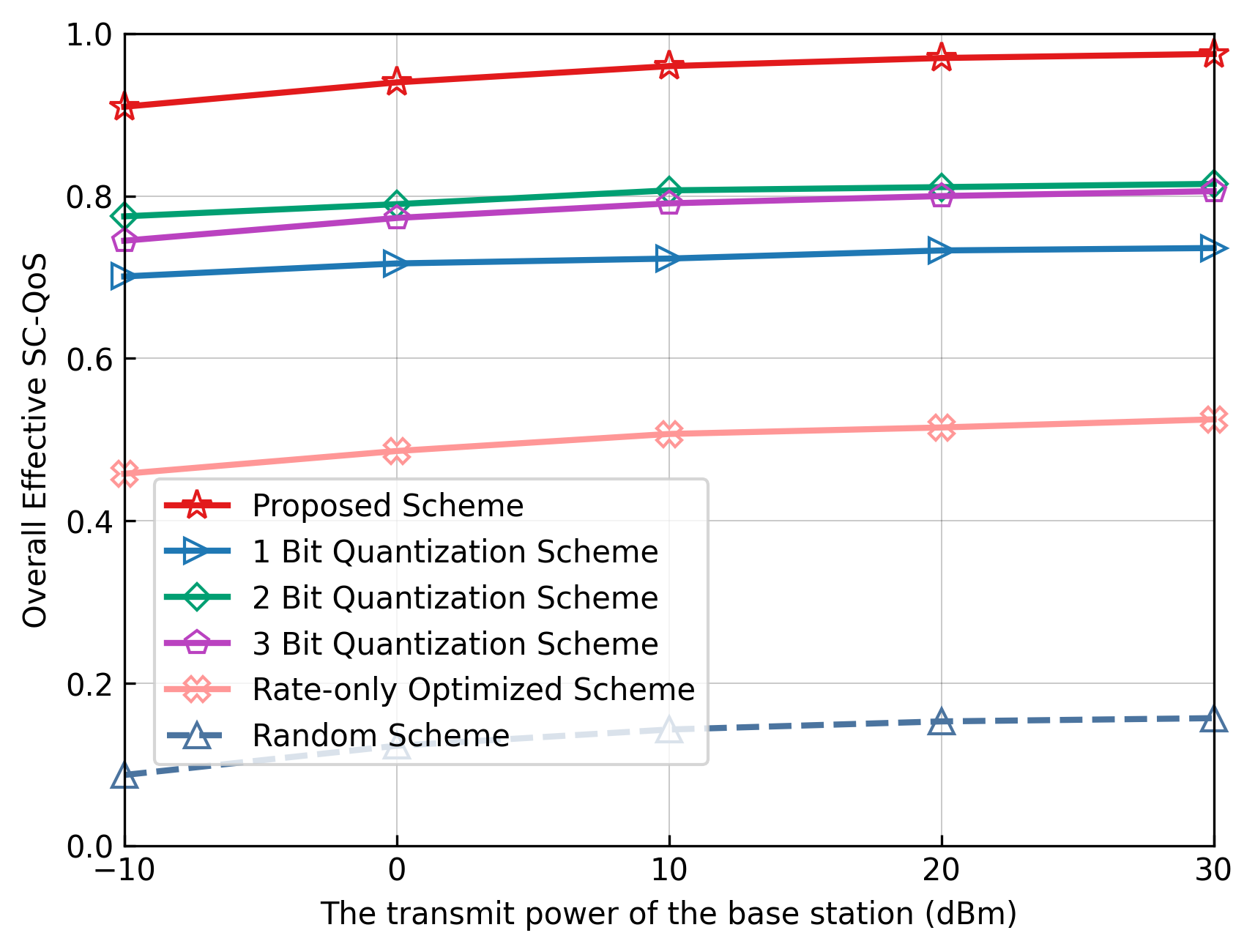}
  \caption{The effective SC-QoS versus the transmit power.}
  \vspace{-0.3cm}
\end{figure}

\begin{figure}
  \centering
  \includegraphics[width=8.2cm]{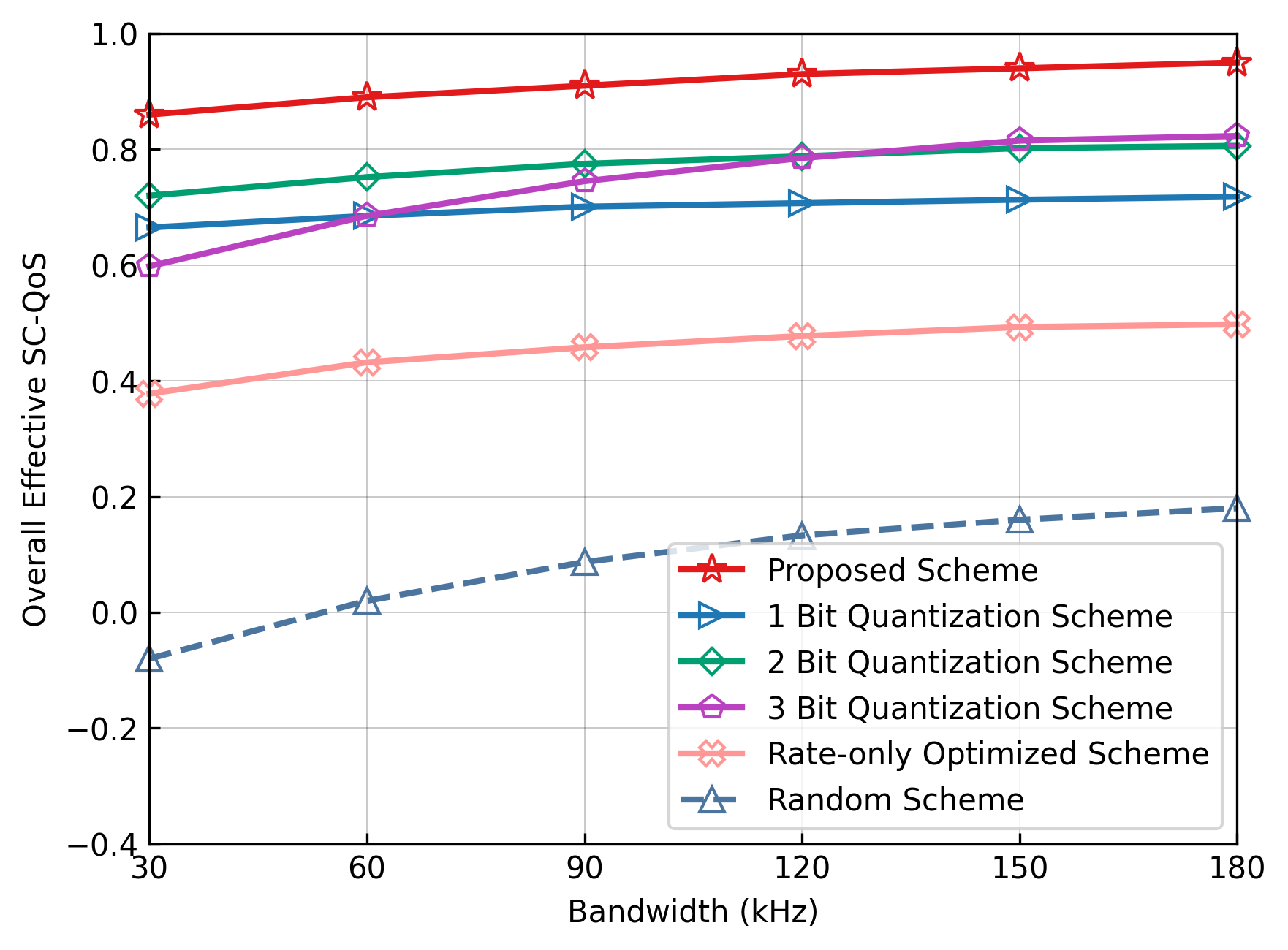}
  \caption{The effective SC-QoS versus the bandwidth.}
  \vspace{-0.3cm}
\end{figure}
The overall effective SC-QoS versus the transmit power and the bandwidth are respectively depicted in Fig. 8 and Fig. 9.
where the ``Rate-only Optimized Scheme'' and ``Random Scheme'' are introduced for better comparison.
In the ``Rate-only Optimized Scheme'', SBQ is performed randomly, while communication resource is allocated by using our proposed scheme.
In the ``Random Scheme'', both SBQ allocation and resource allocation are random.
As observed in Fig. 8 and Fig. 9, the overall effective SC-QoS increases with the transmit power and bandwidth.
This is attributed to the fact that ample communication resources can notably enhance the SNR and transmission rate, leading to a substantial improvement in SC-QoS.
It can be also seen that our proposed scheme overperforms the other benchmark schemes in terms of the SC-QoS.
This is because our proposed intelligent resource allocation scheme is environment-aware and semantic task-aware, and the efficient design of SBQ with compensators can effectively handle the semantic noise.
Futhermore, it is worth noting that the ``Rate-only Optimized Scheme'' and ``Random Scheme'' exhibit weak performance, highlighting the crucial significance of adaptive resource allocation and intelligent strategic decision-making in achieving favorable SC-QoS.


\subsection{SC-QoS wtih Two Resource Allocation Paradigms}
\begin{figure}
  \centering
  \includegraphics[width=8cm]{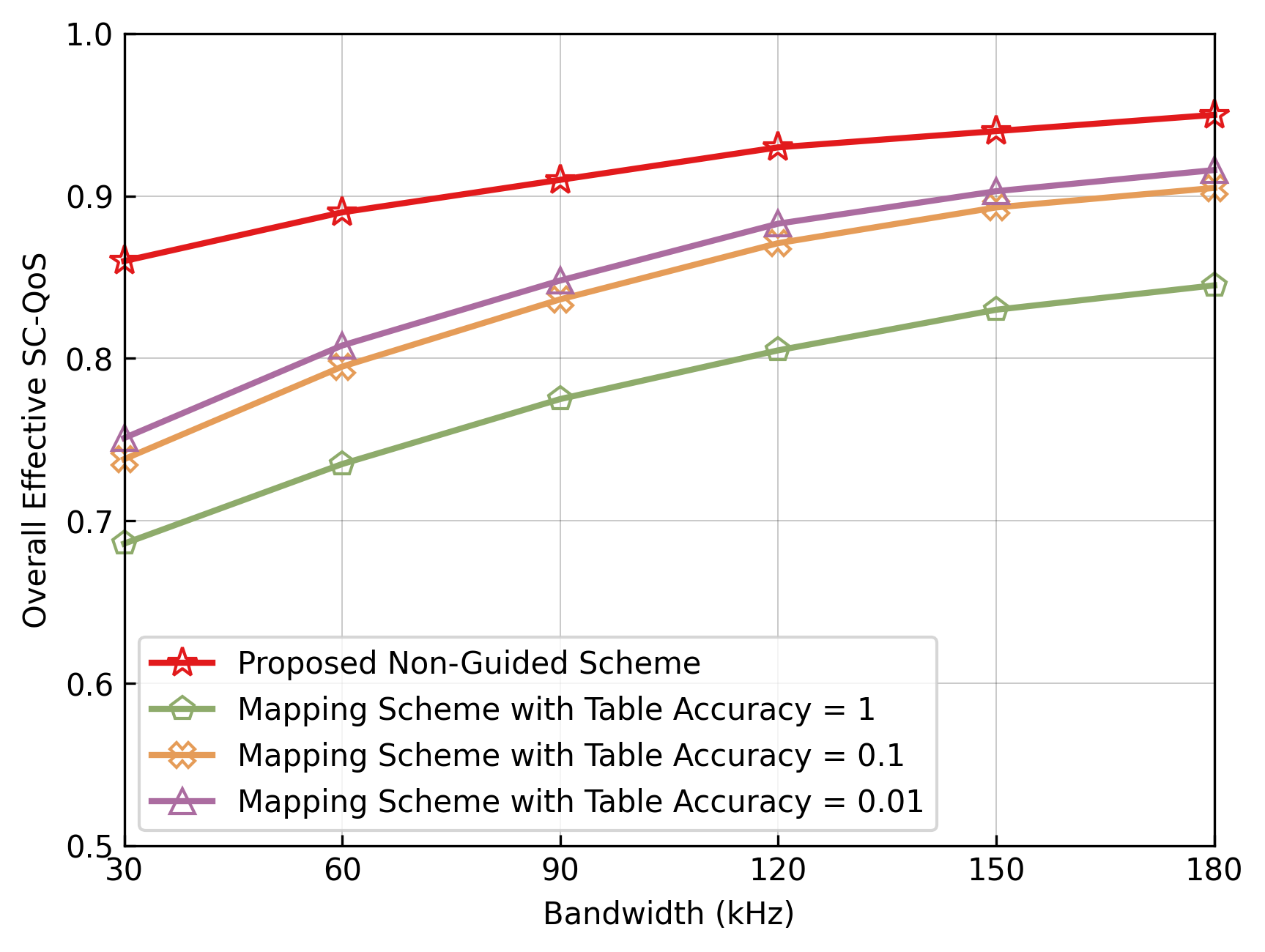}
  \caption{The effective SC-QoS of mapping-guided and non-guided schemes.}
  \vspace{-0.5cm}
\end{figure}
Fig. 10 shows the overall effective SC-QoS of the mapping-guided scheme and the non-guided adaptive scheme for different available bandwidths, where the mapping-guided scheme records semantic similarity with different SNR accuracy. 
As can be observed in Fig. 10, our proposed scheme exhibits better overall effective SC-QoS with $\mathrm{13\%}$ performance improvement. 
This is due to the fact that our non-guided adaptive scheme can eliminate the reliance on mapping guidance and allows the accurate semantic task feedback and the physical channel feedback. 
Conversely, mapping-guided schemes rely on pre-trained guidance for resource allocation, which leads to accuracy loss and limited channel-aware capabilities.
Since narrowband communication is the most common communication method with scarce spectrum, in this case, our proposed non-guided adaptive scheme can also perform better effective SC-QoS than mapping-guided schemes.
Furthermore, it can be seen that the performance of mapping schemes increase with the record accuracy. 
This is because the accuracy loss decreases with increasing record accuracy, at the cost of higher storage costs, especially when the SNR fluctuates in a wide range.

\vspace{-0.1cm} 
\section{Conclusion}
This paper investigated adaptive intelligent resource allocation for semantic communication networks without mapping guidance for the first time.
A novel SBQ approach was proposed to encode the semantic features into bits, and an intelligent offset compensator was considered to effectively eliminate the semantic noise caused by signal distortions and SBQ,  
SC-QoS was defined as a novel performance metric for semantic communication networks based on SQE and semantic transmission. 
The transmit beamforming of the base station, the bits for semantic representation, the bandwidth allocation, and the subchannel assignment were jointly optimized to maximize the effective SC-QoS.
Aiming to address complex non-convex optimization problems involving interdependent variables and advance the development of AI-driven semantic communication networks toward greater intelligence, 
a dynamic intelligent resource allocation scheme by using SAC and D-SAC was designed, which enabled real-time decision-making based on sensed semantic tasks and channel characteristics. 
The simulation results clearly indicated that our proposed adaptive semantic communication networks significantly improved the SC-QoS compared to benchmark schemes,
providing candidate solutions for future ultra-scale intelligent communication networks.

\bibliography{IEEEabrv,reference}

\begin{thebibliography}{10}
\providecommand{\url}[1]{#1}
\csname url@samestyle\endcsname
\providecommand{\newblock}{\relax}
\providecommand{\bibinfo}[2]{#2}
\providecommand{\BIBentrySTDinterwordspacing}{\spaceskip=0pt\relax}
\providecommand{\BIBentryALTinterwordstretchfactor}{4}
\providecommand{\BIBentryALTinterwordspacing}{\spaceskip=\fontdimen2\font plus
\BIBentryALTinterwordstretchfactor\fontdimen3\font minus
  \fontdimen4\font\relax}
\providecommand{\BIBforeignlanguage}[2]{{%
\expandafter\ifx\csname l@#1\endcsname\relax
\typeout{** WARNING: IEEEtran.bst: No hyphenation pattern has been}%
\typeout{** loaded for the language `#1'. Using the pattern for}%
\typeout{** the default language instead.}%
\else
\language=\csname l@#1\endcsname
\fi
#2}}
\providecommand{\BIBdecl}{\relax}
\BIBdecl

\bibitem{LU2020100158}
Y.~Lu and X.~Zheng, ``{6G}: A survey on technologies, scenarios, challenges,
  and the related issues,'' \emph{J. Ind. Inf. Integr.}, vol.~19, pp. 1--52,
  Sep. 2020.

\bibitem{9606720}
K.~B. Letaief, Y.~Shi, J.~Lu, and J.~Lu, ``Edge artificial intelligence for
  {6G}: Vision, enabling technologies, and applications,'' \emph{IEEE J. Sel.
  Areas Commun.}, vol.~40, no.~1, pp. 5--36, Jan. 2022.

\bibitem{8808168}
K.~B. Letaief, W.~Chen, Y.~Shi, J.~Zhang, and Y.-J.~A. Zhang, ``The roadmap to
  {6G}: {AI} empowered wireless networks,'' \emph{IEEE Commun. Mag.}, vol.~57,
  no.~8, pp. 84--90, Aug. 2019.

\bibitem{qin2021semantic}
Z.~Qin, X.~Tao, J.~Lu, W.~Tong, and G.~Y. Li, ``Semantic communications:
  Principles and challenges,'' \emph{arXiv preprint arXiv:2201.01389}, Jun.
  2021.

\bibitem{9530497}
G.~Shi, Y.~Xiao, Y.~Li, and X.~Xie, ``From semantic communication to
  semantic-aware networking: Model, architecture, and open problems,''
  \emph{IEEE Commun. Mag.}, vol.~59, no.~8, pp. 44--50, 2021.

\bibitem{9398576}
H.~Xie, Z.~Qin, G.~Y. Li, and B.-H. Juang, ``Deep learning enabled semantic
  communication systems,'' \emph{IEEE Trans. Signal Process.}, vol.~69, no.~1,
  pp. 2663--2675, Apr. 2021.

\bibitem{9954153}
J.~Xu, B.~Ai, N.~Wang, and W.~Chen, ``Deep joint source-channel coding for
  {CSI} feedback: An end-to-end approach,'' \emph{IEEE J. Sel. Areas Commun.},
  vol.~41, no.~1, pp. 260--273, 2023.

\bibitem{9837870}
T.-Y. Tung and D.~Gündüz, ``{DeepWiVe}: Deep-learning-aided wireless video
  transmission,'' \emph{IEEE J. Sel. Areas Commun.}, vol.~40, no.~9, pp.
  2570--2583, Sep. 2022.

\bibitem{9953076}
D.~Huang, F.~Gao, X.~Tao, Q.~Du, and J.~Lu, ``Toward semantic communications:
  Deep learning-based image semantic coding,'' \emph{IEEE J. Sel. Areas
  Commun.}, vol.~41, no.~1, pp. 55--71, 2023.

\bibitem{9763856}
L.~Yan, Z.~Qin, R.~Zhang, Y.~Li, and G.~Y. Li, ``Resource allocation for text
  semantic communications,'' \emph{IEEE Wireless Commun. Lett.}, vol.~11,
  no.~7, pp. 1394--1398, Jul. 2022.

\bibitem{10122232}
H.~Zhang, H.~Wang, Y.~Li, K.~Long, and A.~Nallanathan, ``{DRL}-driven dynamic
  resource allocation for task-oriented semantic communication,'' \emph{IEEE
  Trans. Commun.}, 2023, to be published.

\bibitem{10101778}
Q.~Hu, G.~Zhang, Z.~Qin, Y.~Cai, G.~Yu, and G.~Y. Li, ``Robust semantic
  communications with masked vq-vae enabled codebook,'' \emph{IEEE Trans.
  Wireless Commun.}, 2023, to be published.

\bibitem{9973061}
W.~Wu, F.~Yang, F.~Zhou, Q.~Wu, and R.~Q. Hu, ``Intelligent resource allocation
  for {IRS}-enhanced {OFDM} communication systems: A hybrid deep reinforcement
  learning approach,'' \emph{IEEE Trans. Wireless Commun.}, 2022, to be
  published.

\bibitem{10107714}
G.~Nan, X.~Liu, X.~Lyu, Q.~Cui, X.~Xu, and P.~Zhang, ``{UDSem}: A unified
  distributed learning framework for semantic communications over wireless
  networks,'' \emph{IEEE Netw.}, 2023, to be published.

\bibitem{wang2023intelligent}
L.~Wang, F.~Yang, Y.~Chen, S.~Lai, and W.~Wu, ``Intelligent resource allocation
  for transmission security on {IRS}-assisted spectrum sharing systems with
  {OFDM},'' \emph{Phys. Commun.}, vol.~58, no. 102013, Jan. 2023.

\bibitem{9832831}
Y.~Wang, M.~Chen, T.~Luo, W.~Saad, D.~Niyato, H.~V. Poor, and S.~Cui,
  ``Performance optimization for semantic communications: An attention-based
  reinforcement learning approach,'' \emph{IEEE J. Sel. Areas Commun.},
  vol.~40, no.~9, pp. 2598--2613, Sep. 2022.

\bibitem{9450827}
Z.~Weng and Z.~Qin, ``Semantic communication systems for speech transmission,''
  \emph{IEEE J. Sel. Areas Commun.}, vol.~39, no.~8, pp. 2434--2444, Aug. 2021.

\bibitem{9796572}
X.~Kang, B.~Song, J.~Guo, Z.~Qin, and F.~R. Yu, ``Task-oriented image
  transmission for scene classification in unmanned aerial systems,''
  \emph{IEEE Trans. Commun.}, vol.~70, no.~8, pp. 5181--5192, Jun. 2022.

\bibitem{9830752}
H.~Xie, Z.~Qin, X.~Tao, and K.~B. Letaief, ``Task-oriented multi-user semantic
  communications,'' \emph{IEEE J. Sel. Areas Commun.}, vol.~40, no.~9, pp.
  2584--2597, 2022.

\bibitem{9959884}
J.~Kang, H.~Du, Z.~Li, Z.~Xiong, S.~Ma, D.~Niyato, and Y.~Li, ``Personalized
  saliency in task-oriented semantic communications: Image transmission and
  performance analysis,'' \emph{IEEE J. Sel. Areas Commun.}, vol.~41, no.~1,
  pp. 186--201, Nov. 2023.

\bibitem{zhang2022unified}
G.~Zhang, Q.~Hu, Z.~Qin, Y.~Cai, G.~Yu, and X.~Tao, ``A unified multi-task
  semantic communication system for multimodal data,'' \emph{arXiv preprint
  arXiv:2209.07689}, Aug. 2022.

\bibitem{10001594}
L.~Yan, Z.~Qin, R.~Zhang, Y.~Li, and G.~Ye~Li, ``{QoE}-aware resource
  allocation for semantic communication networks,'' in \emph{IEEE Glob. Commun.
  Conf. (GLOBECOM)}, Dec. 2022, pp. 3272--3277.

\bibitem{liu2022adaptable}
C.~Liu, C.~Guo, Y.~Yang, and N.~Jiang, ``Adaptable semantic compression and
  resource allocation for task-oriented communications,'' \emph{arXiv preprint
  arXiv:2204.08910}, Apr. 2022.

\bibitem{9031419}
X.~Guan, Q.~Wu, and R.~Zhang, ``Joint power control and passive beamforming in
  {IRS}-assisted spectrum sharing,'' \emph{IEEE Commun. Lett.}, vol.~24, no.~7,
  pp. 1553--1557, Jul. 2020.

\bibitem{9046301}
H.~Zhang, N.~Yang, W.~Huangfu, K.~Long, and V.~C.~M. Leung, ``Power control
  based on deep reinforcement learning for spectrum sharing,'' \emph{IEEE
  Trans. Wireless Commun.}, vol.~19, no.~6, pp. 4209--4219, Jun. 2020.

\bibitem{9381701}
W.~Ahsan, W.~Yi, Z.~Qin, Y.~Liu, and A.~Nallanathan, ``Resource allocation in
  uplink {NOMA-IoT} networks: A reinforcement-learning approach,'' \emph{IEEE
  Trans. Wireless Commun.}, vol.~20, no.~8, pp. 5083--5098, Aug. 2021.

\bibitem{9481307}
A.~M. Seid, G.~O. Boateng, B.~Mareri, G.~Sun, and W.~Jiang, ``Multi-agent {DRL}
  for task offloading and resource allocation in multi-{UAV} enabled {IoT} edge
  network,'' \emph{IEEE Trans. Netw. Service Manage.}, vol.~18, no.~4, pp.
  4531--4547, Dec. 2021.

\bibitem{9435782}
J.~Chen, H.~Xing, Z.~Xiao, L.~Xu, and T.~Tao, ``A {DRL} agent for jointly
  optimizing computation offloading and resource allocation in {MEC},''
  \emph{IEEE Internet Things J.}, vol.~8, no.~24, pp. 17\,508--17\,524, Dec.
  2021.

\bibitem{9759989}
B.~Hazarika, K.~Singh, S.~Biswas, and C.-P. Li, ``{DRL}-based resource
  allocation for computation offloading in {IoV} networks,'' \emph{IEEE Trans.
  Industr. Inform.}, vol.~18, no.~11, pp. 8027--8038, Nov. 2022.

\bibitem{vaswani2017attention}
A.~Vaswani, N.~Shazeer, N.~Parmar, J.~Uszkoreit, L.~Jones, A.~N. Gomez,
  {\L}.~Kaiser, and I.~Polosukhin, ``Attention is all you need,'' in
  \emph{Proc. Adv. Neural Inf. Process. Syst.}, Dec. 2017, pp. 5998--6008.

\bibitem{devlin2018bert}
J.~Devlin, M.-W. Chang, K.~Lee, and K.~Toutanova, ``Bert: Pre-training of deep
  bidirectional transformers for language understanding,'' \emph{arXiv preprint
  arXiv:1810.04805}, 2018.

\bibitem{lan2019albert}
Z.~Lan, M.~Chen, S.~Goodman, K.~Gimpel, P.~Sharma, and R.~Soricut, ``{ALBERT}:
  A lite {BERT} for self-supervised learning of language representations,'' in
  \emph{Proc. Int. Conf. Learn. Representation (ICLR)}, Addis Ababa, Ethiopia,
  Apr. 2019, pp. 1--17.

\bibitem{yang2019xlnet}
Z.~Yang, Z.~Dai, Y.~Yang, J.~Carbonell, R.~Salakhutdinov, and Q.~V. Le,
  ``{XLNet}: Generalized autoregressive pretraining for language
  understanding,'' in \emph{Proc. Adv. Neural Inf. Process. Syst. (NeurIPS)},
  Vancouver, Canada, Dec. 2019, pp. 5754--5764.

\bibitem{liu2019roberta}
Y.~Liu, M.~Ott, N.~Goyal, J.~Du, M.~Joshi, D.~Chen, O.~Levy, M.~Lewis,
  L.~Zettlemoyer, and V.~Stoyanov, ``{Roberta}: A robustly optimized {bert}
  pretraining approach,'' \emph{arXiv preprint arXiv:1907.11692}, Jul. 2019.

\bibitem{sac2}
T.~Haarnoja, A.~Zhou, K.~Hartikainen, G.~Tucker, S.~Ha, J.~Tan, V.~Kumar,
  H.~Zhu, A.~Gupta, P.~Abbeel \emph{et~al.}, ``Soft actor-critic algorithms and
  applications,'' \emph{arXiv preprint arXiv:1812.05905}, 2018.

\end{thebibliography}

\end{document}